\documentclass[10pt,journal]{IEEEtran}

\usepackage{moreverb}

\usepackage{graphicx}
\usepackage[usenames,dvipsnames]{xcolor}
\usepackage{overpic}

\usepackage{amsmath}
\usepackage{float} 
\usepackage{caption}
\usepackage{subcaption}
\usepackage{algorithm}
\usepackage{algpseudocode}

\newcommand{\beq}{\begin{eqnarray}}
\newcommand{\eeq}{\end{eqnarray}}

\newcommand{\require}{\textbf{Input: }}
\newcommand{\ensure}{\textbf{Output: }}
\usepackage[absolute,overlay]{textpos} 

\begin{document}

\title{Wireless Gas Leak Detection and Localization}
\author{Fabien~Chraim, Yusuf~Bugra~Erol, Kris~Pister\\
\IEEEauthorblockA{Berkeley Sensor and Actuator Center, University of California, Berkeley, USA\\ \{chraim, yberol, pister\}@eecs.berkeley.edu}
   \IEEEcompsocitemizethanks{
      \IEEEcompsocthanksitem F.~Chraim and K.~Pister are with the Berkeley Sensor and Actuator Center (BSAC), University of California, Berkeley, CA, USA.
      \IEEEcompsocthanksitem Yusuf~Bugra~Erol is with the Center for Information Technology Research in the Interest of Society (CITRIS), University of California, Berkeley, CA, USA.
      \IEEEcompsocthanksitem Copyright (c) 2015 IEEE. Personal use of this material is permitted. However, permission to use this material for any other purposes must be obtained from the IEEE by sending a request to pubs-permissions@ieee.org
  }
}

\maketitle

\begin{textblock*}{0.9\textwidth}(0.12\textwidth,1.1\textheight)
\centering
\footnotesize
© 2015 IEEE. This is the author's accepted manuscript of an article published in IEEE. The final authenticated version is available online at https://doi.org/10.1109/TII.2015.2397879.
\end{textblock*}

\markboth{Wireless Gas Leak Detection and Localization}{}

\begin{abstract}
Thousands of industrial gas leaks occur every year, with many leading to injuries, deaths, equipment damage, and a disastrous environmental effect. There have been many attempts at solving this problem, but with limited success. This paper proposes a wireless gas leak detection and localization solution. With a monitoring network of 20 wireless devices covering 200m$^2$, 60 propane releases are performed. The detection and localization algorithms proposed here are applied to the collected concentration data, and the methodology is evaluated. A detection rate of 91\% is achieved, with seven false alarms recorded over three days, and an average detection delay of 108 seconds. The localization results show an accuracy of 5 meters. Recommendations for future explosive gas sensor design are then presented.
\end{abstract}

\section{Introduction}
\label{intro}
The number of gas leaks that occur every year on industrial plants is unknown. Most of these leaks, even if detected, go unreported when they don't directly lead to tangible accidents. Environmental Protection Agency (EPA) reports estimate that, in the United States alone, these plants emit close to one billion cubic meters of methane (not taking any other gas into consideration). Most of these losses (around 80\%) seem to come from leaky compressors, valves, seals and connectors \cite{EPA2013lessons}. In 2012, approximately 2,200 million metric tons of $CO_2$ equivalent were accidentally released from petroleum systems and other chemical processes necessary for the production of plastics, cement, iron and steel \cite{EPA-web}. It is estimated that around 800,000 to 900,000 leaks are investigated each year on refineries, with between 200 and 300 of them having directly resulted in loss of life, injuries, damaged equipment, or operational losses \cite{fact-sheet}. In short, industrial gas leaks present a major challenge in the quest for safe, environmentally-friendly, and cost-effective plants.

In this paper, we present a distributed wireless sensor approach to the problem of gas leaks in large industrial spaces (chemical plants, refineries, oil rigs, etc.). The objective is to detect and localize ``refinery-like'' gas leaks within seconds of their occurrence. With many corporations upgrading their facilities with a low-power wireless infrastructure (WirelessHART) \cite{greg2010}, a leak detection system that simply connects to the wireless umbrella would be a desired addition to the existing safety framework. Our goal is to study the feasibility of such an approach, while carefully reviewing some of the detection challenges in the hope for opening the door for widespread commercial adoption.

The remainder of this article is organized as follows. In section \ref{sec:lit-review} we will review some of the available solutions to the problem at hand, both at the academic level and commercially. Our approach is then presented in section \ref{sec:approach}, where we look at the system architecture, the hardware and the detection/localization algorithms. Our experimental results are shown in section \ref{sec:experimental}, in which we validate our approach using real Propane leaks. Future directions and recommendations are left for section \ref{sec:conclusion} where we conclude.

\section{Literature Review and State of the Art}
\label{sec:lit-review}
The review presented in this section is divided into a survey of commercial methods for leak detection, and some of the ideas coming out of academia.

\subsection{Gas Leak Detection Systems}
Conventional leak detection methods fall under two categories: fixed instrumentation and mobile sensing. In the former, a sensor is affixed in the general vicinity of equipment suspected of leaking (valves, compressors, ...). These instruments are usually connected to a constant power source and generate alarms based on their sampled data. These alarms can be visual or audible, or can feed directly into a plant management system. Mobile sensors are usually hand-held devices that a worker has to point at the suspected leak source and evaluate the readings on the spot. Reports of the measurements are relayed in real-time either through a wireless connection, or by direct communication between the worker and other plant employees. Both these methods have their advantages and disadvantages and most often, a hybrid system of fixed and mobile sensors is implemented. In particular, a fixed sensor is able to continuously monitor an area, as opposed to a worker who samples the same region for a few seconds perhaps before moving on. Fixed sensors have better instruments by virtue of the fact that they are less constrained, but mobile sensors allow the operator to trace a leak to its origin. It is obvious that mobile sensors put the worker at risk during the sampling process, while the fixed sensors enable safer operation \cite{fixedmobile}.

In this study, we are only interested in fixed instruments because our proposed solution is static in nature. We now look at some of the commercially available solutions for comparison. Many solutions have been proposed for the problem of leak detection in pipelines\cite{survey2012}. This topic, though relevant, is not of direct interest here since leak detection near pipelines can be accomplished by deploying a series of sensors in a linear sequence. The solution presented in this article would not be very practical for long pipeline installations due to the high number of sensors it would require.

Perhaps, the most prevalent leak detection methodology is by concentration measurement. Pellistor, electrochemical, semiconductor and infrared sensors are all used to sample the ambient gas for particular species. By means of preset threshold detection, alarms are raised alerting workers and plant operators \cite{sensor-survey2012}. These widely adopted sensors however suffer from one or more of the following: low sensitivity, short lifetime, high energy consumption, sensitivity to ambient conditions, high costs, drift... Typically, these sensors are operated independently, meaning that no information about the source of the leak is given. Due to the fact that they consume a considerable amount of energy, installing them becomes an issue as virtually always, the cost of laying cables outstrips the cost of the device itself\cite{sensor-survey2012}.

Pipeline diagnosis systems gave rise to ultrasonic sensors which have recently been adopted in some plants. The principle of operation relies on the fact that gas leaks sometimes come from punctured pipes which emit acoustic ``tone'' signals in the ultrasonic band \cite{emerson-acoustic}. These sensors, though unaffected by environmental conditions, do not measure the intensity of the leak, and are still unable to determine its origin. They are designed to work with gases under pressure, and do not represent a general solution to industrial gas leaks.

In recent years, camera systems have found their way to the gas leak detection and localization market. These devices tend to be mounted on elevated towers, often rotating to cover the entirety of the plant. They operate by taking snapshots of the environment, then analyzing the sampled images to detect gas leaks \cite{rebellion}. Most of the available solutions on the market operate in the infrared band, but recently more versatile snapshot hyper-spectral instruments are being utilized.

Energy and cost present challenges for large-scale deployments of gas sensors. Some studies have focused on improving the sensing methodology to address these issues \cite{laserspeck, deployment}. So et al. present an optically-based gas detector for various species. Their solution is based on commercial off-the-shelf components, and employs photo-acoustic spectroscopy making it tunable to various species. Their results are promising as they achieve a reduction in power consumption as well as a reduced manufacturing cost. They validate their sensor experimentally with $CO_2$, measuring concentrations down to 410 ppb. Their designs demonstrate a \$2,000 device with current consumption on the order of 70mA \cite{laserspeck}.

Somov et al. \cite{deployment} develop a hazardous gas detection system based on wireless battery-powered devices. Their methane sensor is a planar catalytic one built on gamma alumina membranes. Their circuitry achieves a reduced power consumption. Their boiler room deployment consists of nine wireless methane sensors and a gateway. The average power consumption of their device is at 2.64 mW, with a sampling interval of 30 seconds and a transmission every five minutes. As a result, the lifetime of their device is at 641 days. Their application targets gas sensing resolutions of 0.15\% volume of methane.

\subsection{Detection and Localization Using Multiple Sensors}
Academically, the problem of detecting and localizing leaks has been addressed in many fields. Under different names, similar methodologies have been applied to detecting the location of a speaker using many microphones, localizing objects using multiple radar streams, etc. We now list of few relevant examples.

Glenn et al. propose using inverse diffusion modeling to localize leaks. By assuming a fickian diffusion model, they consider a large network of sensors surrounding the source of the leak. Coupling diffusion with Ensemble Kalman filtering allows them to estimate the location of the source of the leaks. Their simulated system reports plume origins as numerous hypotheses each having likelihoods \cite{Nofsinger2004diffusion}.

Huseynov et al.\cite{Huseynov2009mems} propose a distributed network of MEMS ultrasonic sensors for gas leak localization. In their study, a comparison of energy-decay (ED) and time-difference of arrival (TDOA) methods for localization is presented. With a distributed network of four devices, they attempt to localize a nitrogen leak from a small orifice. They employ maximum likelihood (ML) and the least squares (LS) techniques to find closed form solutions for the diffusion differential equations. In their deployment, a 20 ft × 20 ft room is instrumented with 4 MEMS microphones (running at 200 kHz). Nitrogen gas was released at 150 psi at four different locations. They successfully localize the nitrogen leak with an accuracy lower than 1 ft.

Weimer et al. consider gas leaks in wide and dense wireless sensor networks. The problem being addressed is one of large scale leaks, with high concentration of gases (such as harmful gases in a metropole). In their model therefore, they take diffusion and air currents into account. An interesting idea is presented concerning the subsampling of sensors which are in close proximity, in order to reduce the network-wide energy consumption. A wake-up process ensures that all the devices are running when needed. Their method combines binary hypothesis testing with Kalman filtering, and is implemented on a testbed of 30 wireless light sensors \cite{Weimer2009wsn}.

The methods presented in these articles (and more on the topic of gas leak detection and localization) have merit as they present valid and interesting theoretical ideas and simulations. However, they all leave much to be desired in the space of experimental validation. In this study, we attempt to solve the problem of gas leak detection and localization in the most applied manner possible. We employ some mathematical and statistical tools, and apply them to real gas concentration measurements recorded during a series of intentional releases performed in a typical industrial setting.

\section{A Wireless Distributed Sensing Approach}
\label{sec:approach}
In this paper, we study the problem of gas leak detection and localization by means of a wireless, distributed network of sensors. Though such an approach could be viewed as a standalone system of sensors, it could also benefit from further integration. The gas sensors utilized here could be tacked on to other industrial process control instruments. For example, the wireless valve positioning solution presented in \cite{chraim-valve}, could be augmented with a gas leak sensor, as both of these problems are often linked. Furthermore, a wireless perimeter security solution \cite{chraim-fence} could also assist in detecting and localizing plant leaks as soon as suspicious concentrations leave the enclave of the plant. In the remainder of the paper, we consider the leak detection solution as a stand-alone system.

\subsection{Hardware}
\begin{figure}
  \centering
    \includegraphics[width=0.5\textwidth]{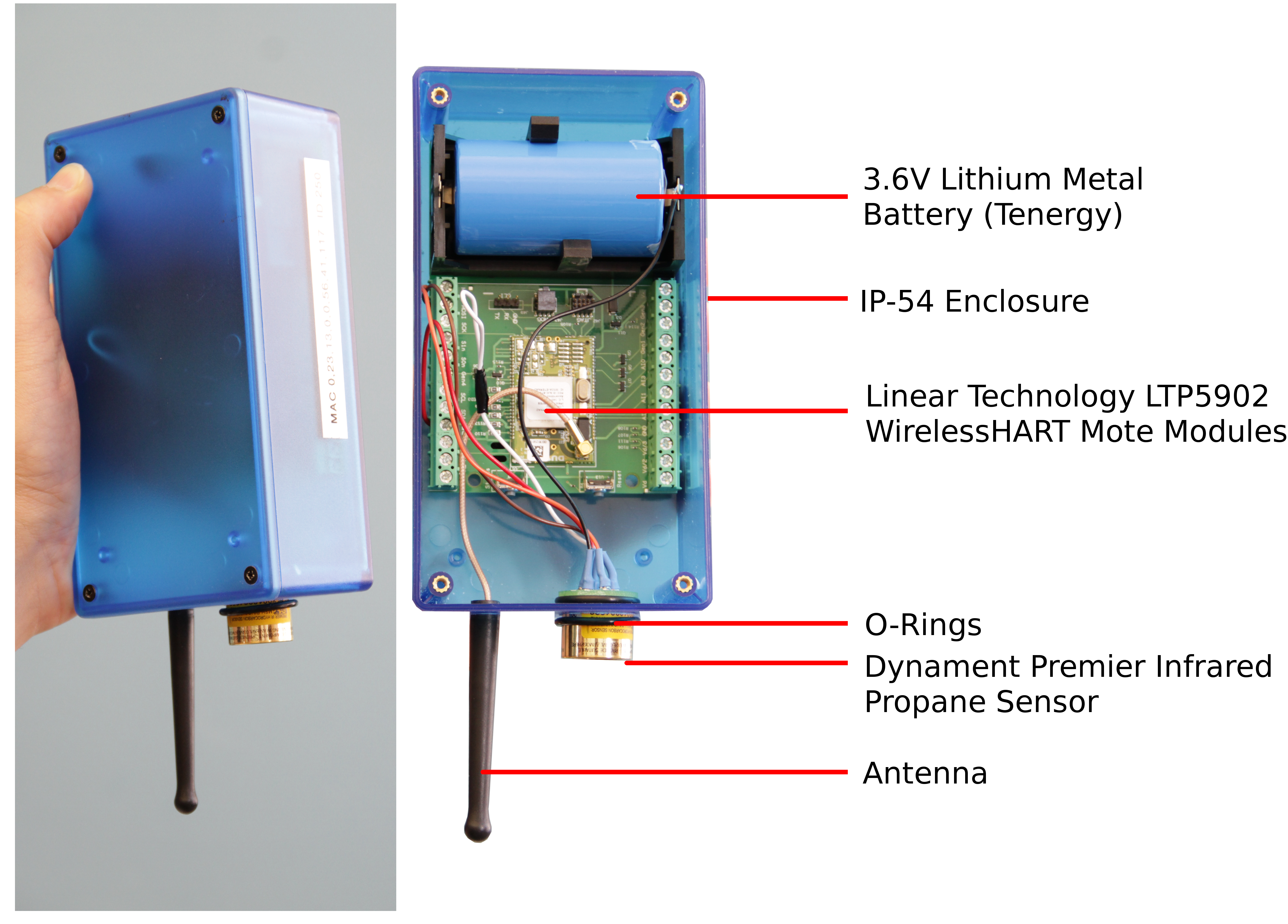}
  \caption{The hardware platform used in this study: a wireless, battery-powered propane sensor.}
  \label{fig:mote}
\end{figure}

\begin{figure}
\centering
   \begin{subfigure}{1\linewidth}
     \includegraphics[scale=0.6]{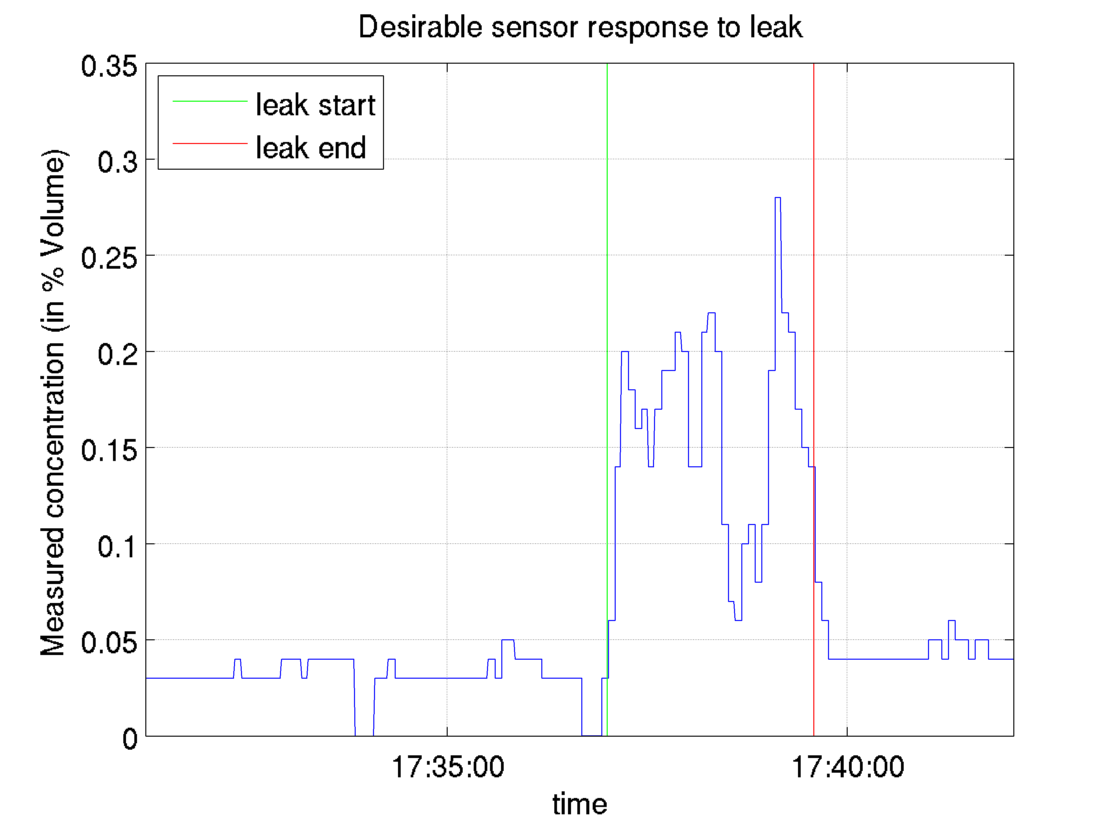}
     \caption{Sensor 1 response: even with a high noise floor, this sensor was responsive to the leak.}
   \end{subfigure}
   \\
   \begin{subfigure}{1\linewidth}
     \includegraphics[scale=0.6]{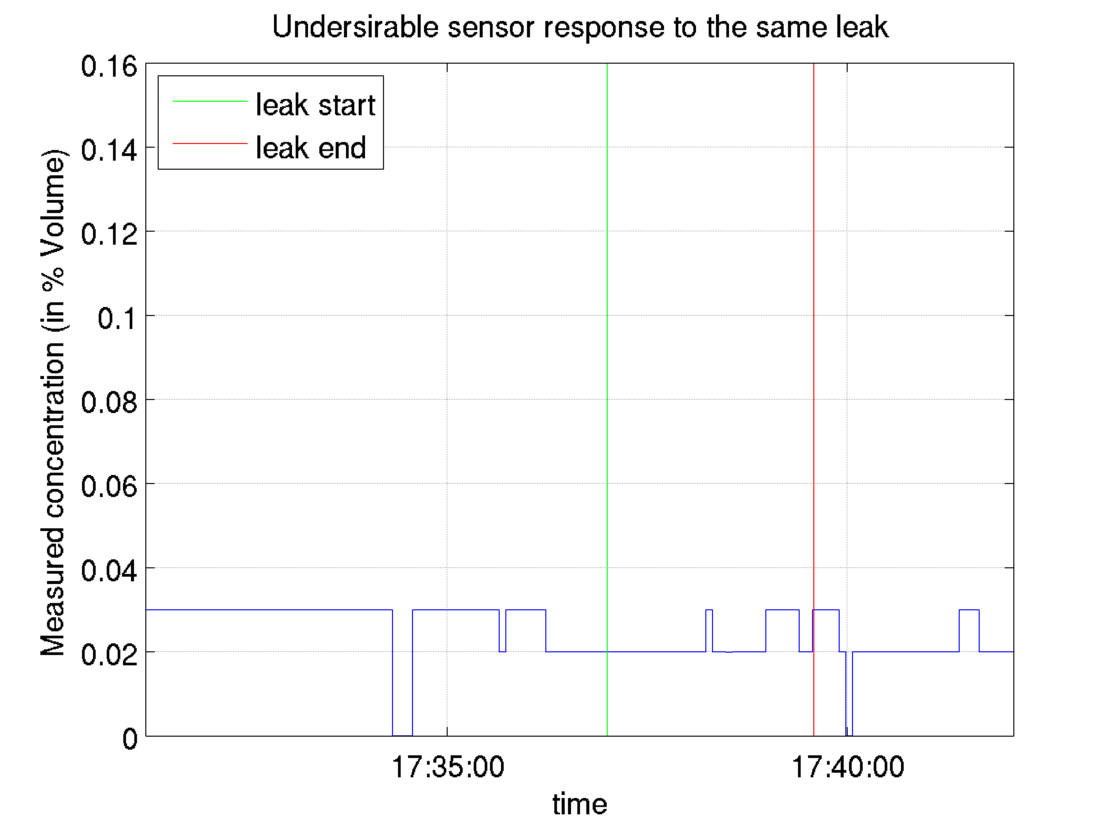}
     \caption{Sensor 2 response: elevated noise and no apparent response to the same leak.}
   \end{subfigure}
    \\
   \begin{subfigure}{1\linewidth}
     \includegraphics[scale=0.15]{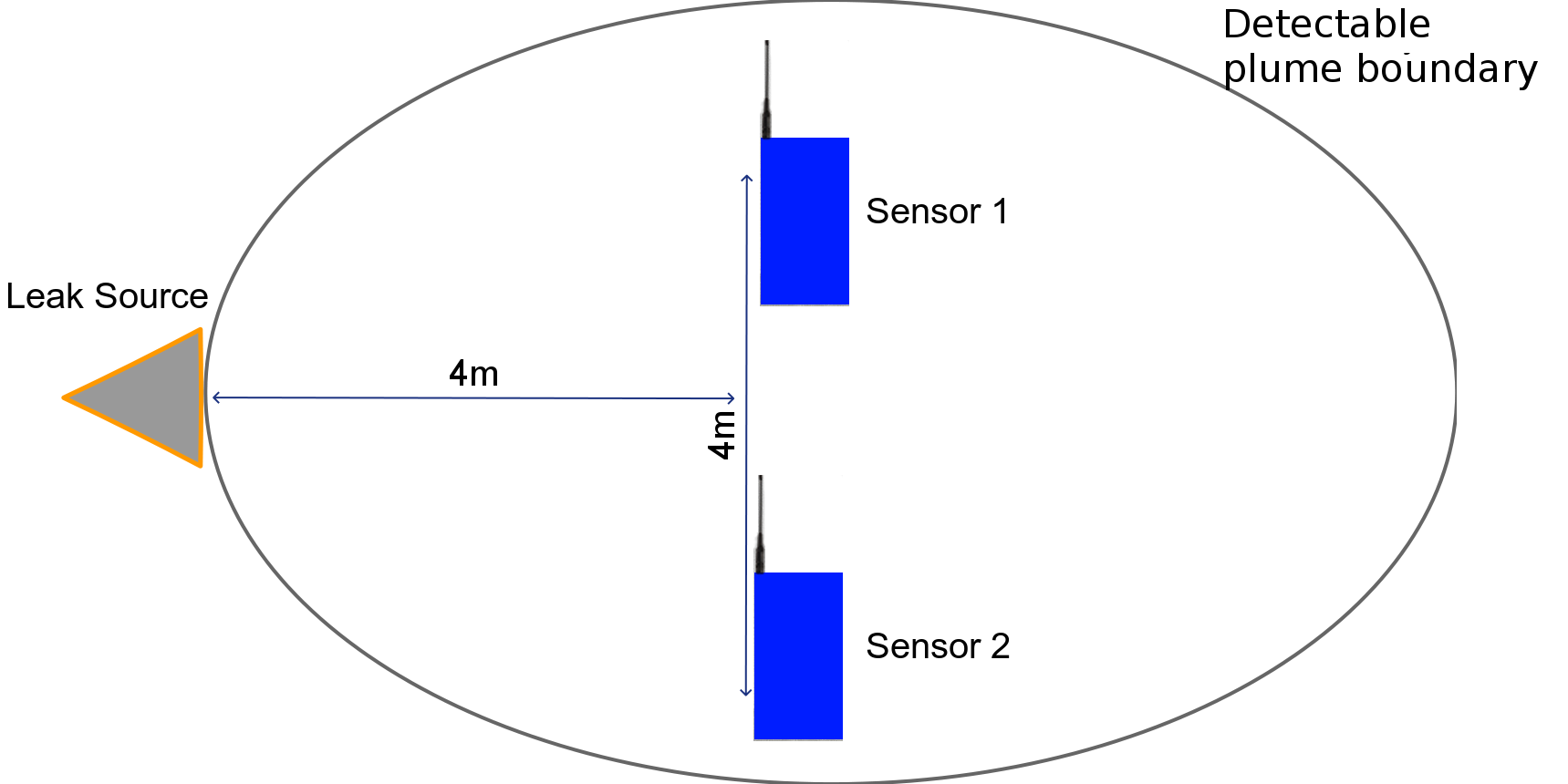}
     \caption{Experimental setup diagram showing both sensors within the plume boundary during the leak.}
   \end{subfigure}
 \caption{Sensor SNR and Response challenges: in this experiment, two sensors were positioned 4m apart, and 4m away from the source of the leak. These sensors were observed with a FLIR Camera and validated to be present in a detectable plume.}
  \label{fig:goodbad}
\end{figure}

As is customary with most wireless sensing applications, the hardware we developed in this study includes a radio, a microcontroller, a sensor and a power source. Recently, the push for higher integration in the industry resulted in system-on-chip (SoC) solutions for the microcontroller and radio. In this project, the Linear Technology LTP5902 SmartMesh WirelessHART Mote Modules were utilized. They feature a 32 bit ARM Cortex M3 microcontroller, along with a 2.4 GHz, IEEE 802.15.4, WirelessHART (IEC62591) compliant radio. Combining the time-synchronized channel hopping protocol with extremely low transmission and reception power levels (on the order of 5mA at 3V), these modules achieve a 99.999\% network reliability with a sub-50$\mu$A average currents.\\
In this project, the focus is on explosive gases. For this reason, all of the studies and releases were performed around propane, which is a by-product of natural gas processing and appears in the process of refining petroleum. This gas is representative of the family of explosive gases, and was made available to us for use in the various experimental stages. Propane is commonly used in commercial and residential applications, and is in the liquefied petroleum (LP) gases family. As such, a propane sensor needed to be integrated with the LTP5902 SoC. The Dynament Premier Infrared Hydrocarbon (propane) Sensor (MSH-P/HC/NC/5/V/P) was selected, with a 0-2\% volume measurement range (or 0 - 20,000 part per million (ppm)) and a resolution of 0.01\% volume (100 ppm). Our integrated device is shown in figure \ref{fig:mote}. Though this propane sensor is, at the time of writing, one of the best on the market, its performance leaves a lot of room for improvement. For starters, it consumes an average of 80mA of current (at 3V), and possesses a start-up time of 1 minute. This certainly represents a challenge in battery operated devices. Additionally, the sensor has a temperature compensation routine to account for changes in the ambient temperature. However, this feature does not have a very fast response time. This means that in outdoor installations especially, small gusts of wind which lead to temperature changes can result in large variations in the concentration measurement (by hundreds of ppms). Although the sensor features an internal temperature reading, this measurement itself does not represent the ambient temperature around the housing of the sensor. Rather, this reading is largely affected by the heating elements inside the sensor. Shown in figure \ref{fig:goodbad}, is the response of two identical sensors deployed outdoors in the vicinity of a leak source. It is clear that state-of-the-art propane sensors still present many challenges, and we hope to demonstrate that they are the missing link in the productization of this solution.\\
Considering the average power levels of the communication module, and with careful application design and network configuration, it is often possible to implement wireless sensing applications where lifetimes extend beyond the shelf life of batteries (around 10 years). However, in this wireless gas leak application, the sensor remains as the limiting factor and a burden on the energy budget. 
Finally, it is worth noting that the hardware was powered by an industrial D-size Lithium metal battery, with 19 Ah of charge (with a duty-cycling of 25\%, this battery would hold enough charge for about 40 days). The device was enclosed in an ABS plastic IP-54 enclosure by Hammond.

\subsection{System Architecture}
\begin{figure}
  \centering
    \includegraphics[width=0.5\textwidth]{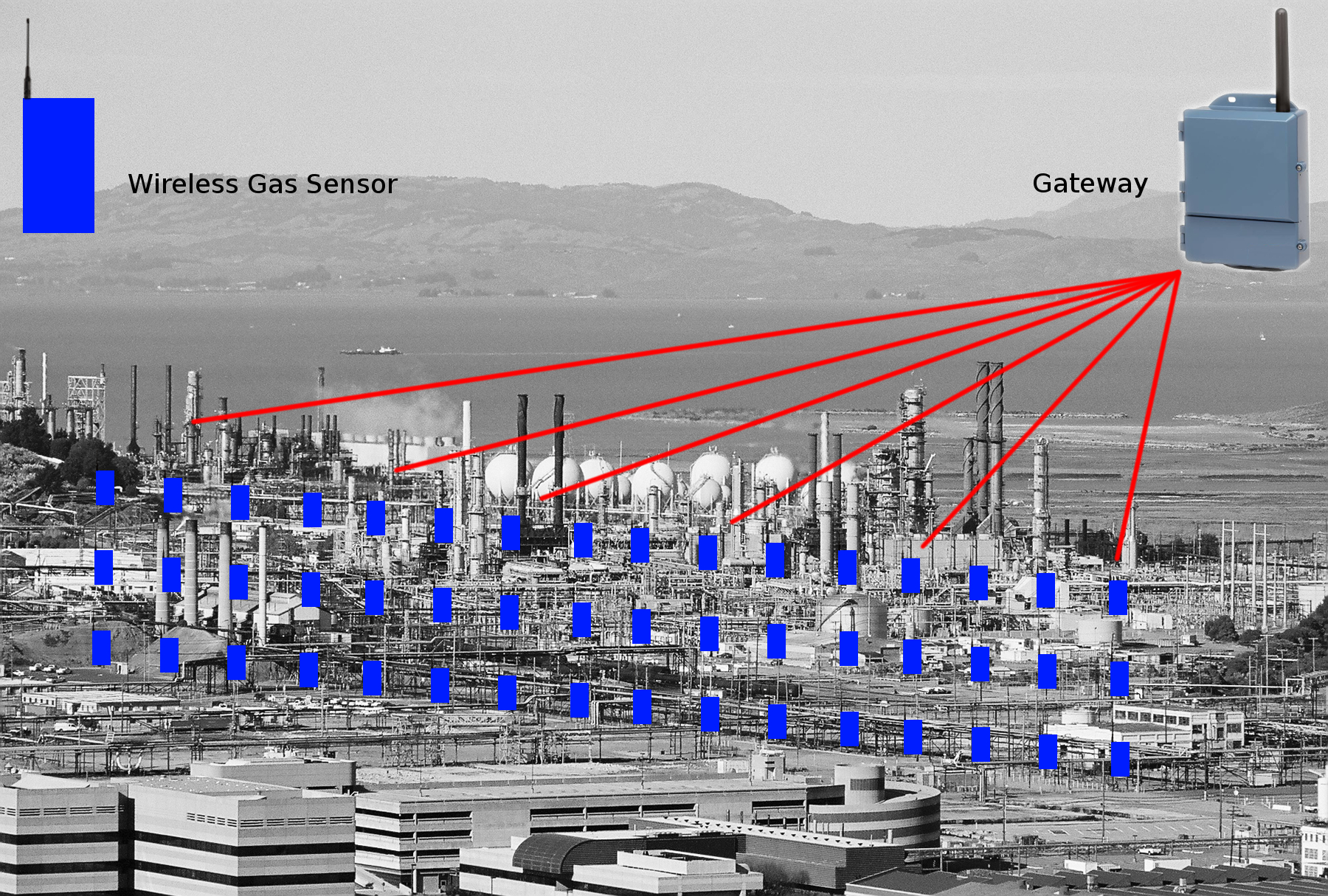}
  \caption{Proposed system architecture: gas leak detection sensors are deployed extensively across a sensitive industrial area (a refinery in this case); data travels through the mesh network towards a single collection point (gateway) where the detection and localization algorithms are applied. The sensors can be duty-cycled spatially and temporally based on the measured concentrations.}
  \label{fig:architecture}
\end{figure}

The system architecture for the leak detection solution is perhaps best explained with a picture (see figure \ref{fig:architecture}). Taking a refinery for the sake of example, gas leak detection sensors would be deployed throughout a refinery. Though a grid distribution is often easier to manage, it is not a required feature. Indeed, these easy to install sensors can be mounted in seconds to the existing buildings and poles, and as long as their location is recorded, the concentration data will be easily processed by the algorithms presented later. The path from data to decision starts with concentration measurements at the node side, which are filtered and transmitted to the gateway (when needed), using the wireless infrastructure. The gateway algorithms then collect the concentration data from the many sensor devices on the grounds and generate alarms whenever a leak is detected. Additionally, periodic reports concerning the concentration gradients of gases on the refinery can be generated and sent to concerned parties.\\

In terms of sensor placement, these devices would be spread throughout the plant. However, it is beneficial to increase the density in zones susceptible to leaks. For example, an increase in the number of compressors and valves raises the risk of leaks. As such, a typical deployment would have a minimum density necessary for detections, and then certain areas would be characterized with clusters of sensors (increased density) based on the risk. To minimize the overall application energy consumption (for the entire system of sensors), neighboring devices might take turn in ``guarding'' a zone, then alerting the other devices in case of suspicious increases in measured concentrations.

Similar to having an adaptive spatial sampling of concentration, a temporal one would also be beneficial. This means that the mote can decide to increase its sampling frequency when it detects a sudden surge in gas concentration. The reporting rate of data to the gateway, which is not very frequent regularly (on the order of one average reading every 10 minutes for example) can also be increased when the device records an unusual concentration of gas. Augmenting this method with a statistical routine can also get rid of many unnecessary alarms and minimize energy consumption. The experiments performed in this study were designed with oversampling. In long-term deployments however, this large amount of data would not be necessary. Instead, sensor would transmit steady-state concentrations to the gateway in the form of averages and other statistical figures.

As for the gateway, common practices involve powering it directly from the mains. However, it would be possible to utilize a solar scavenger with a rechargeable battery instead. Combined with a low-power Linux box, the WirelessHART gateway will consume on the order of 1.5 Watts on average.

\subsection{Detection Algorithm}
\begin{figure}
\centering
   \begin{subfigure}{1\linewidth}
     \includegraphics[scale=0.25]{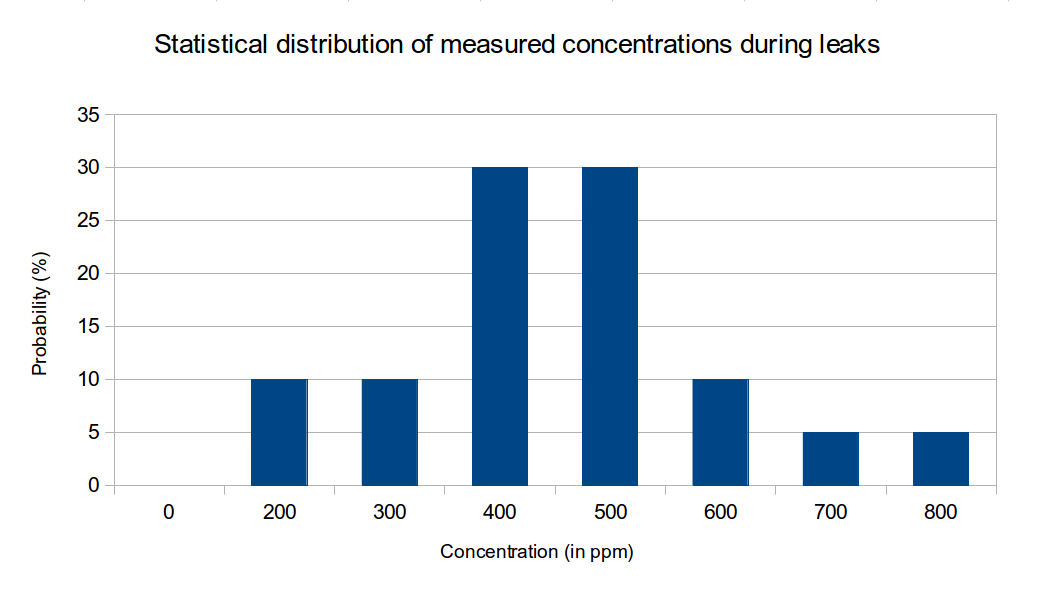}
     \caption{Sensor model for concentration measurments during a leak (ON Model).}
   \end{subfigure}
   \\
   \begin{subfigure}{1\linewidth}
     \includegraphics[scale=0.25]{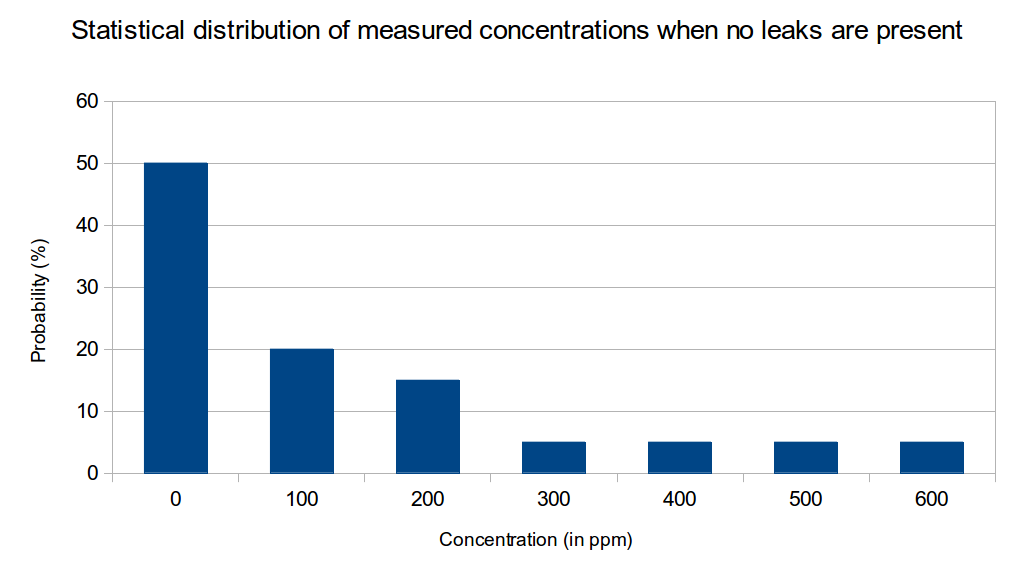}
     \caption{Sensor model for concentration measurments during a leak (OFF Model).}
   \end{subfigure}
 \caption{Semiheuristic sensor model derived from experimental data. Non-complementary probabilities accompany each state: ON (leak occurring) or OFF (no leak). This model was obtained by observing various sensor behaviors during leaks and in their absence (similar to the one shown in figure \ref{fig:goodbad}). The concentration counts were then performed and adjusted heuristically as shown in these histograms. As will be apparent, our detection method is based on the variations of probabilities in a particular period of time.}
  \label{fig:on_off_model}
\end{figure}

Considering the system architecture above, where each sensor adaptively reports its gas concentration measurements to one location (through one or more gateways), we now look at a method for detecting the occurrences of leaks. Our framework is a probabilistic one, where each sensory observation $s(t)$ is represented probabilistically, then the total likelihood of a leak is computed at every time step. To get there, we model each sensor observation independently as $p(s_i(t) \mid \theta_t)$ where $i \in \left\{1,\dots,M\right\}$, $M$ is the total number of sensors per area under consideration, and $\theta_t$ is an indicator variable which represents the existence of a leak at time step $t$. We have derived semiheuristic models for our sensors both in the presence of a leak (ON, $\theta =1$) and when just measuring leak-free environments (OFF, $\theta =0$). The measurements and experiments leading to these models will be defined in section \ref{sec:experimental}. The histograms corresponding to the derived models are illustrated in Figure \ref{fig:on_off_model}.

\begin{figure}
  \centering
    \includegraphics[width=0.5\textwidth]{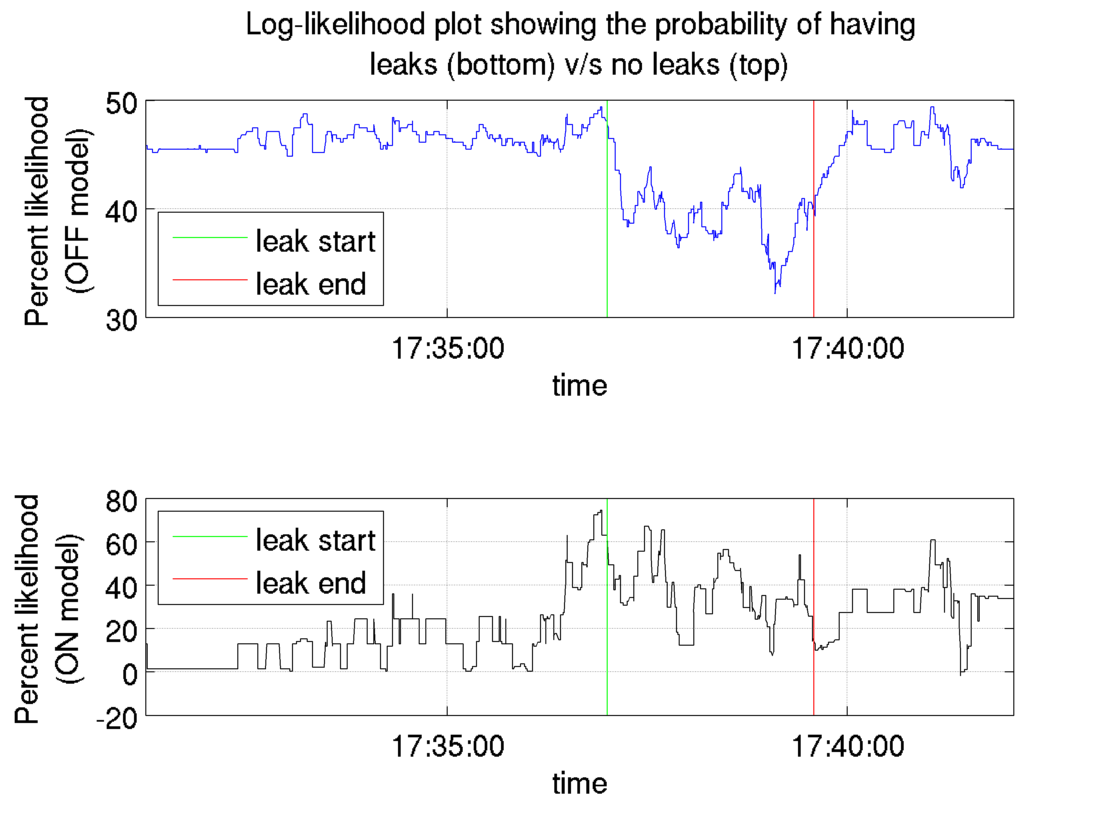}
  \caption{Observation likelihood at each time step, for both states: as concentration measurements are received, the probability of having a leak v/s no leak are computed. In the event of a gas release, one would expect to see the likelihood of having no leaks drop, while that of having leaks increase.}
  \label{fig:likelihood}
\end{figure}

Using these models, the likelihood at each time step is computed as follows.
\beq
	L_t(\theta) = \prod_{i=1}^N p(s_i(t) \mid \theta)
\eeq
When $\theta =1$, we are essentially computing the likelihood of a leak being present (ON state). $L_t(\theta = 0)$ is the likelihood in the OFF state, or when no leaks are present. This is done at each time step, using all of the available sensory information $s_i(t)$ for $i \in \left\{1,\dots,M \right\}$. Intuitively, one would expect $L_t(\theta = 1)$ to increase in the presence of a leak (i.e. ON state), while expecting $L_t(\theta = 0)$ to decrease during the same time. Such a behavior is observed in our experiments and figure \ref{fig:likelihood} is one sample case.

In the quest for a completely automatic leak detection system, we would like to record these changes in the likelihood signal autonomously. We therefore treat the problem as a signal segmentation one \cite{lavielle1998optimal}. In essence, we are interested in the segments of the likelihood timeseries which differ from the normal. Such signal segmentation techniques have been successfully applied in ECG (electrocardiogram) and EEG (electroencephalogram) applications \cite{varon2012robust,mico2010automatic}. We are using the algorithm presented in \cite{varon2012robust} to detect changes in our constructed likelihood signal $L_t(\theta)$, in both states. The method is detailed in algorithm \ref{alg:change}. The Auto-Correlation Functions (ACFs) of segments of length $N$ are computed, as this windows progresses along the timeseries. Each ACF computation is called $A_i$, where $i \in \left\{1,\dots,L/N\right\}$, $L$ is the total length of the timeseries. The cosine of pair-wise ACF computations ($A_i, A_j \mid i \neq j$) is performed, forming a similarity matrix of size $L/N \times L/N$. The weights of the similarity matrix are then calculated and indices outside of the $P$-th percentile are labeled.

The main intuition behind this is the following: disturbance free segments of a signal should have \emph{similar} ACFs. However, if there is a clear change in the signal, there should be a corresponding \emph{dissimilarity} in the ACFs as well. Computing the pairwise cosine similarities between each ACF one can visualize the similarity between the segments. By summing the columns of the resultant similarity matrix, the weights are computed. In our application, when there are no leaks, these weights are high. During a leak however, we expect the weights to decrease as they will be \emph{dissimilar} to the regular ``no-leak'' segments. The change in the segments is then detected simply by applying a threshold set at the $P$-th percentile value.

\begin{center}

\begin{algorithm}
\caption{Phase-1 change detection algorithm}\label{alg:change}
\begin{algorithmic}[1]

\\
\require likelihood timeseries, window size $N$, percentile value $P$\\
\ensure indices of the segments corresponding to the irregularities\\

Segment the signal into epochs of length $N$ \; \\
Compute the Auto-Correlation Functions (ACFs) of each segment, $A_i$ \; \\
Compute the cosine similarity between each pair of ACFs 
\beq
	\cos \sigma_{ij} = \frac{A_i^T A_j}{\left\| A_i \right\| \left\| A_j \right\|}
\eeq	
and form the similarity matrix\; \\
Compute the weights by summing the columns of the similarity matrix\; \\
Calculate the $P$-th percentile of the weights\; \\
Label the segments with weights outside the $P$-th percentile\; 

\end{algorithmic}
\end{algorithm}

\end{center}

\begin{figure}
  \centering
    \includegraphics[width=0.5\textwidth]{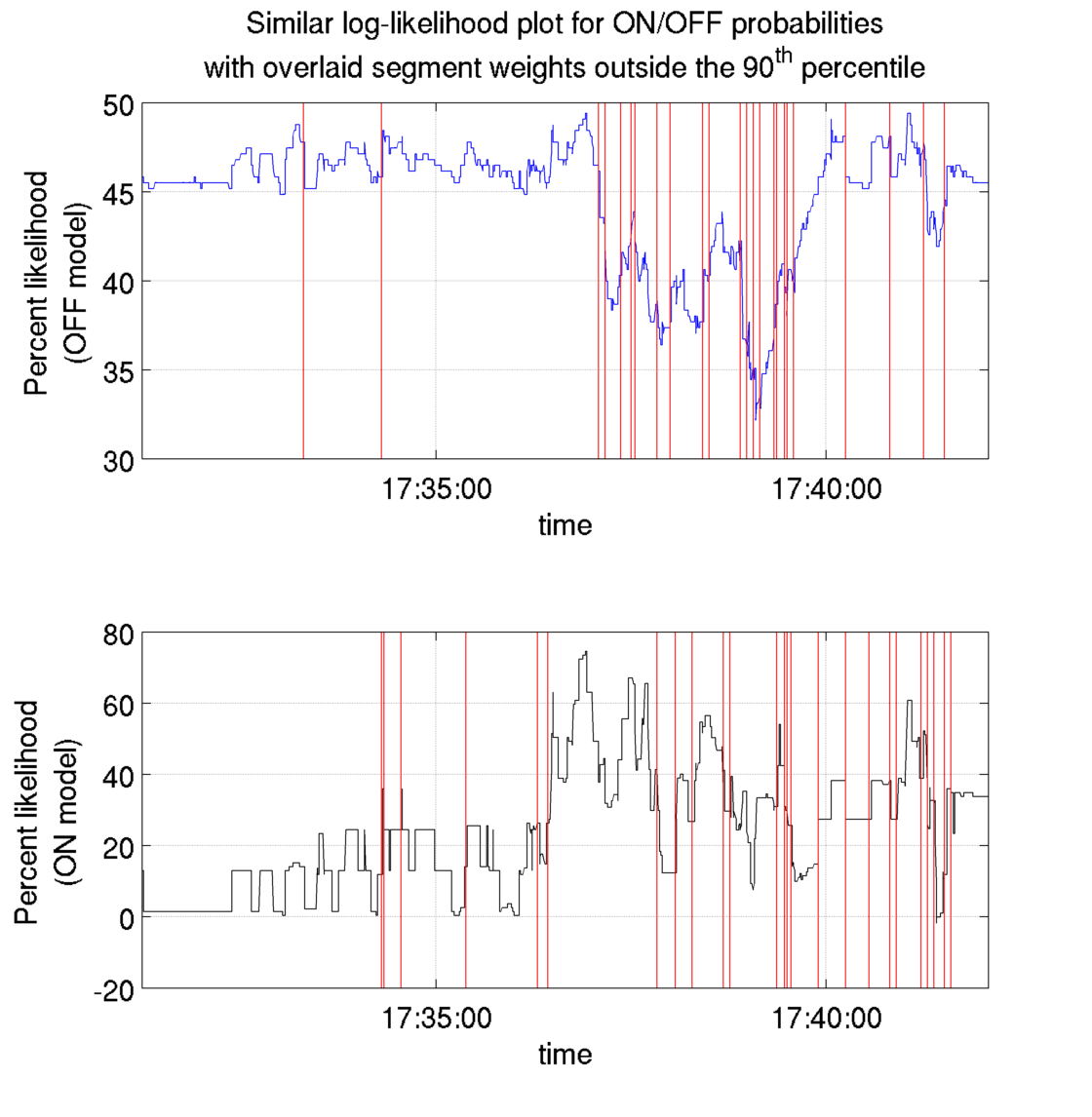}
  \caption{The algorithm described here takes the likelihood timeseries as input and returns the instants in time where that plot went beyond the norm (by a certain percentile threshold).}
  \label{fig:segmentation}
\end{figure}

The algorithm presented here then returns a number of time indices which correspond to the various regions of the likelihood series, which were unusual. We will name these time indices as stage-1 detections. Figure \ref{fig:segmentation} shows the same likelihood plots of figure \ref{fig:likelihood}, but with overlayed segments representing the stage-1 detections.

\begin{figure}
  \centering
    \includegraphics[width=0.5\textwidth]{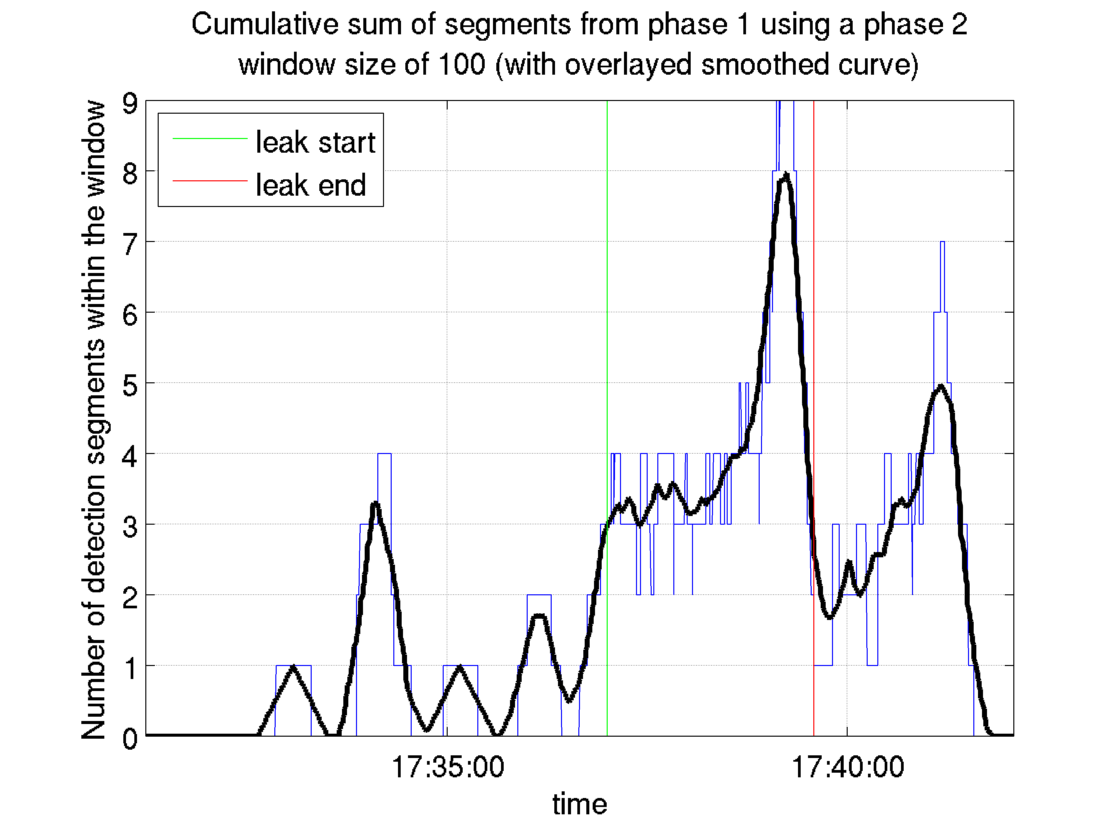}
  \caption{Summing the detections in stage-2 allow for an easier identification of leaks. Depending on the window size, the amount of detections can cross a detection threshold for a considerable time. Shown here is the summation plot, overlayed with a smoothed version for clarity.}
  \label{fig:loglikelicount}
\end{figure}

One of the important parameters of this algorithm is the percentile parameter. Depending on the percentile threshold which is set, one can reduce the rate of false positives. These false positives arise from the fact that the sensor noise floors are elevated (as explained in part A), and therefore reducing our signal-to-noise ratio. Increasing the percentile threshold in our algorithm reduces the false alarm rate, but this however comes at the price of an reduction in true positives. This parameter should be therefore set in such a way that \emph{alarm fatigue} and a ``tolerable'' leak miss rate are balanced.

Another effect of the sensor noise is the fact that some of the stage-1 detections actually correspond to fluctuations of concentration measurements triggered by temperature variations around the sensing element. This explanation was obtained from the sensor manufacturer. This effect means that we are not able to immediately consider the output of the first stage and have to perform additional steps before making a decision. To reduce \emph{false positives}, we look at the total number of stage-1 detections in a sliding window. A sample output of stage-2 is shown in figure \ref{fig:loglikelicount}, where a window size of 100 was utilized. If this cumulative count of detections exceeds some pre-determined threshold level for a preset period of time, we then output a detection, which we name as a stage-2 detection. The length of the sliding window, the threshold level, and the duration of crossing, can all be used to control the false positive and true positive rates, similarly to stage-1. The overall performance of stage-2 detections (and therefore that of the entire algorithm) is analyzed with respect to user-set parameters (percentile threshold, stage-1 window size, and stage-2 window size). The results are presented in section \ref{sec:experimental}.

\subsection{Localization Algorithm}
\begin{figure}
\centering
   \begin{subfigure}{1\linewidth}
   \centering
    \includegraphics[scale=0.55]{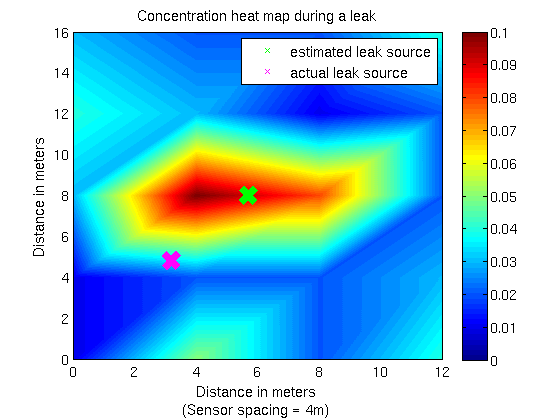}
     \caption{Concentration heat map during a leak. The sidebar denotes concentrations in \% volume.}
   \end{subfigure}
   \\
   \begin{subfigure}{1\linewidth}
   \centering
     \includegraphics[scale=0.55]{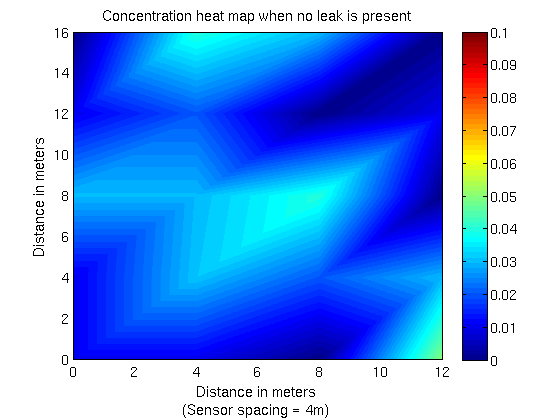}
     \caption{Baseline concentration heat map (no gas is present). The sidebar denotes concentrations in \% volume.}
   \end{subfigure}
 \caption{Concentration heat maps of a 5x4 grid of sensors in two states: during a leak and when no gas is present.}
  \label{fig:localization}
\end{figure}

Following a successful detection of a leak by both stages of the algorithm, a localization routine is called. We utilized a simple center of mass approach. Upon finding a detection, we calculate the 2-dimensional mean of the concentration measurements in the in the X-Y plane.
\beq
    \hat{x} &= \frac{\sum\limits_{i=1}^N s_i(t)x_i}{\sum\limits_{i=1}^N s_i(t)}\\
    \hat{y} &= \frac{\sum\limits_{i=1}^N s_i(t)y_i}{\sum\limits_{i=1}^N s_i(t)}
\eeq	
In the above equations, $x_i$ and $y_i$ represent the coordinates of each sensor, while $s_i(t)$ is the sensor concentration reading. The resulting point $(\hat{x} , \hat{y})$ is defined as the estimate of the leak source detection. This would correspond to finding the point of maximum concentration on a heat map. Figure \ref{fig:localization} depicts the localization result on a particular heat map.

\subsection{Possible Improvements to our Method}
\label{sec:alg_improvements}
In our analysis, we assumed that the sensory readings are conditioned only on the state of the leak (i.e. ON vs OFF). Now, as the deployment grows in size, a far away sensor from the leak source will probably not be able to detect any change. Still, considering a reduced spatial sampling, the response of the sensor to the plume of gas will depend on its location with respect to this plume. To take this into account, the sensor models would be augmented in the following form.
\beq
	p(s_i(t) \mid \theta, Location )
\eeq
However, developing sensor models based on plumes is well beyond the scope of this work, and it would somehow complicate the study. Nevertheless, it is worth mentioning that a location-dependent model can be worked out similarly to the previous section. The likelihood methods will then be directly applied, with some tuning.

One other important improvement that can be introduced is time dependence. With our current approach, we do not exploit the fact that if a leak exists at time $t$, it is then likely for this leak to remain at time $t+1$. Considering this feature means that we could use a general state space model to describe the leak phenomenon. Our objective would then be to perform state estimation, for which one may resort to Kalman filtering, Unscented Kalman Filtering or Particle Filtering, etc.

\section{Experimental Validation}
\label{sec:experimental}
\begin{figure}
  \centering
    \includegraphics[width=0.5\textwidth]{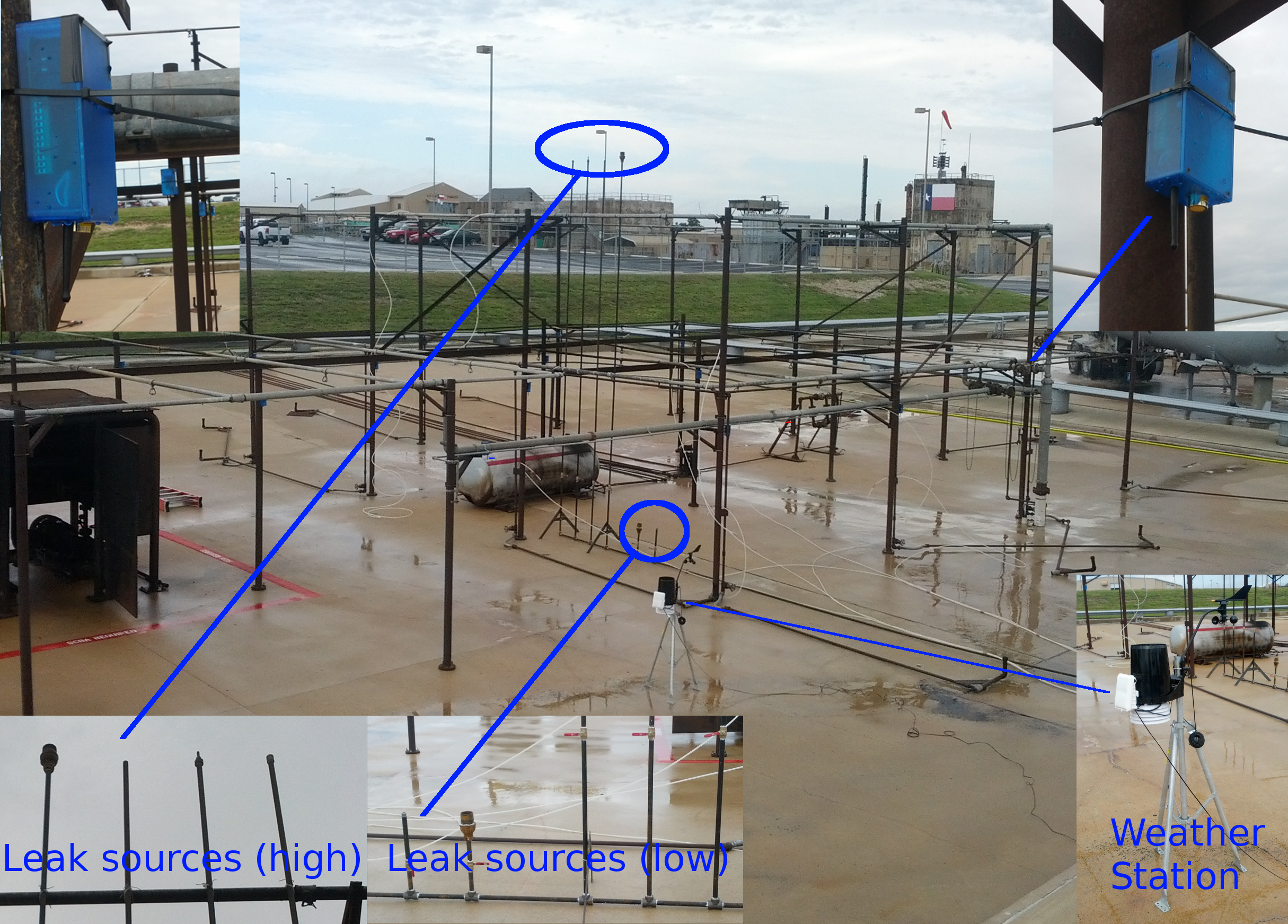}
  \caption{Site of the experiment in College Station, TX. This figure shows the two release points, and the placement of the 20 sensor grid (5x4) at an elevation of about 2.25m.}
  \label{fig:site}
\end{figure}

To validate our architecture, hardware and algorithms, we took part in an experiment at the Texas A\&M Engineering Extension Service facility, in College Station, TX. Over the period of three days, more than 60 propane leaks of two minutes each were released. These leaks were monitored using the detection system presented here, and controlled by a team of engineers and a team of firefighters. The site of the releases is shown in figure \ref{fig:site}. Twenty wireless propane sensors were used to monitor an area of about $200m^2$ surrounding the two release points (at $0.5m$ and $5.5m$). The sensors were placed in a $4 \times 5$ grid configuration, with a separation of about $4m$. All of them were mounted on an elevation of about $2.25m$. These devices measured propane concentrations at a rate of 1 measurement every 5 seconds. The measurements were collected in a data packet and transmitted to a nearby gateway. We now present the results of our algorithm applied to the 60 releases of different source heights, source nozzle sizes ($2mm$, $6.35mm$, $19mm$, and $63.5mm$), and flow rates (ranging between $1.35 lb/hr$ and $1020 lb/hr$).

\subsection{Detection Results}
\begin{figure}
\centering
   \begin{subfigure}{1\linewidth}
   \centering
    \includegraphics[scale=0.55]{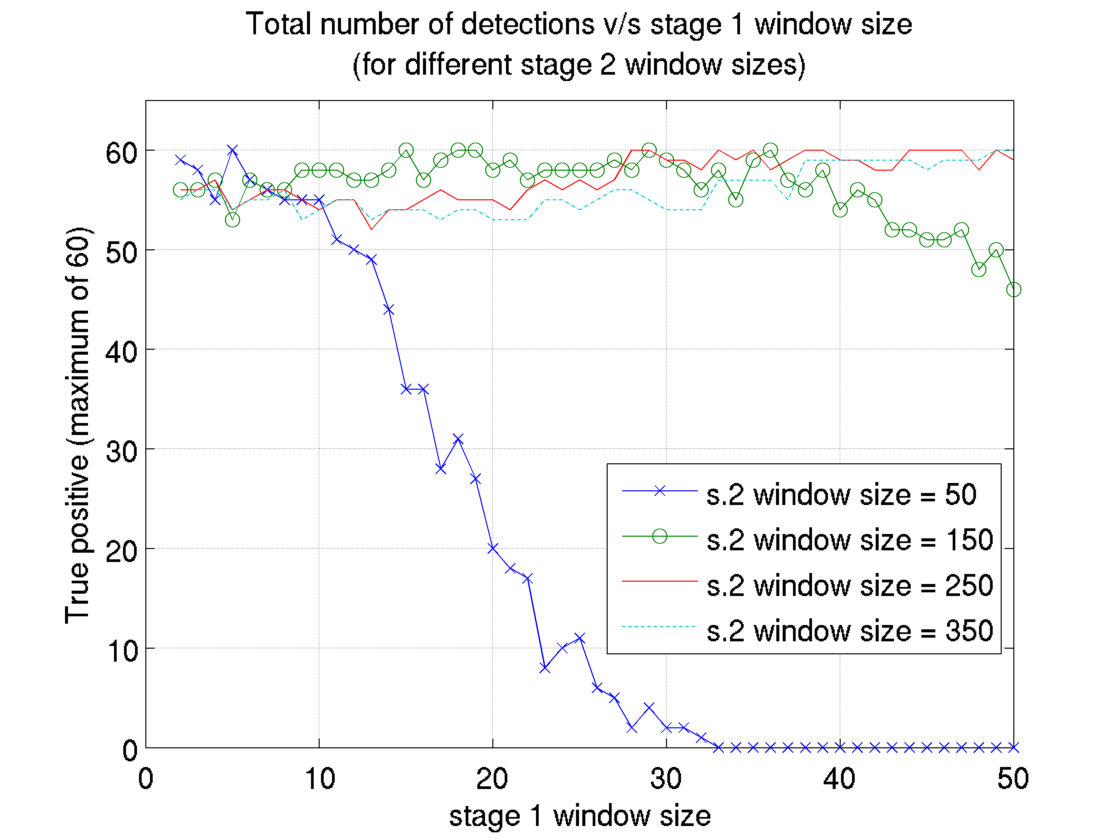}
     \caption{An increase in the window size generally leads to a decrease in the number of detections.}
   \end{subfigure}
   \\
   \begin{subfigure}{1\linewidth}
   \centering
     \includegraphics[scale=0.55]{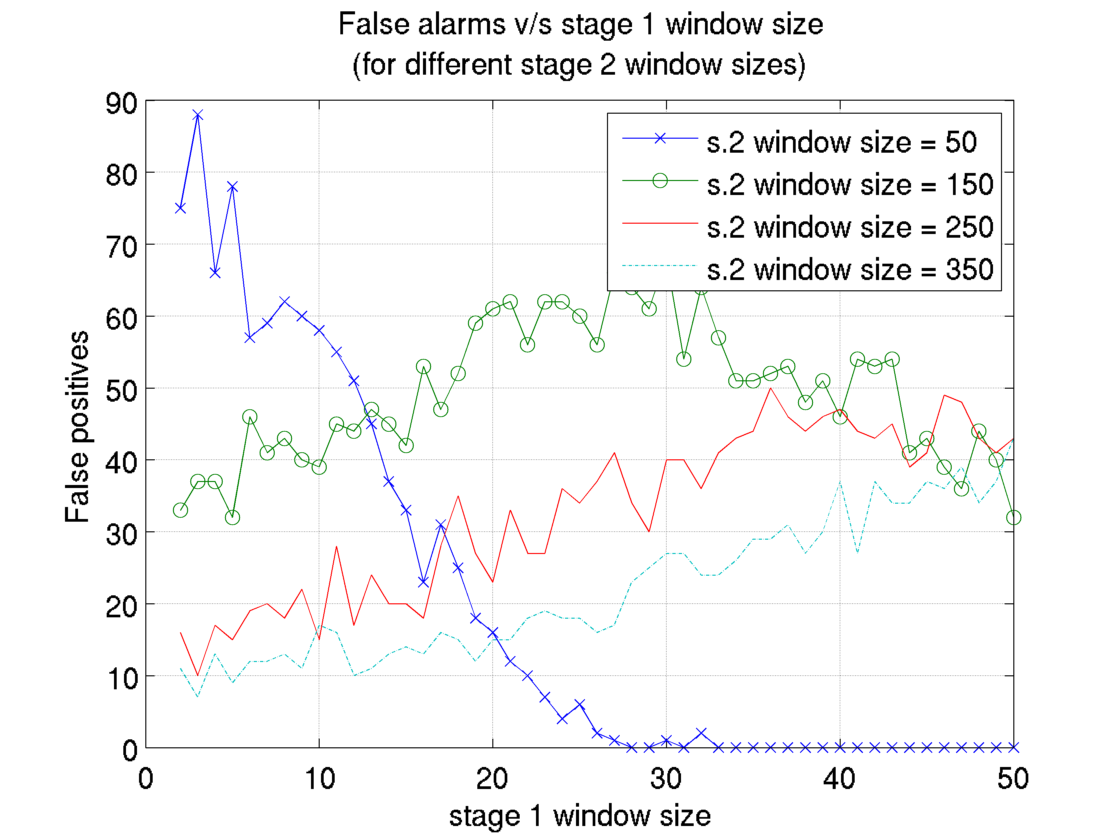}
     \caption{False alarms also increase with the window size.}
   \end{subfigure}
 \caption{Impact of varying the stage-1 window size on the number of detections and the false alarms.}
  \label{fig:det-false-p1}
\end{figure}

The algorithms described in section \ref{sec:approach} were applied to the collected concentration measurements, with different parameters modified for every pass. First we look at the number of correct detections and false alarms as the stage-1 window size is varied. The results are shown in figure \ref{fig:det-false-p1}. The general trend observed shows that increasing the window size of the first stage does not affect the number of detections greatly, except when it grows to a point where the actual variations associated with the leaks are no longer beyond the selected percentile threshold. This effect is accelerated when the window size of stage-2 is reduced. The effect in question is readily visible in figure \ref{fig:det-false-p1} (a) for the stage-2 window size of 50. Looking at false positives, we notice that the trend peaks at a particular stage-1 window size before starting to roll off. However, it is to note that this roll-off is sometimes associated with a reduction in detections as well.

\begin{figure}
\centering
   \begin{subfigure}{1\linewidth}
   \centering
    \includegraphics[scale=0.55]{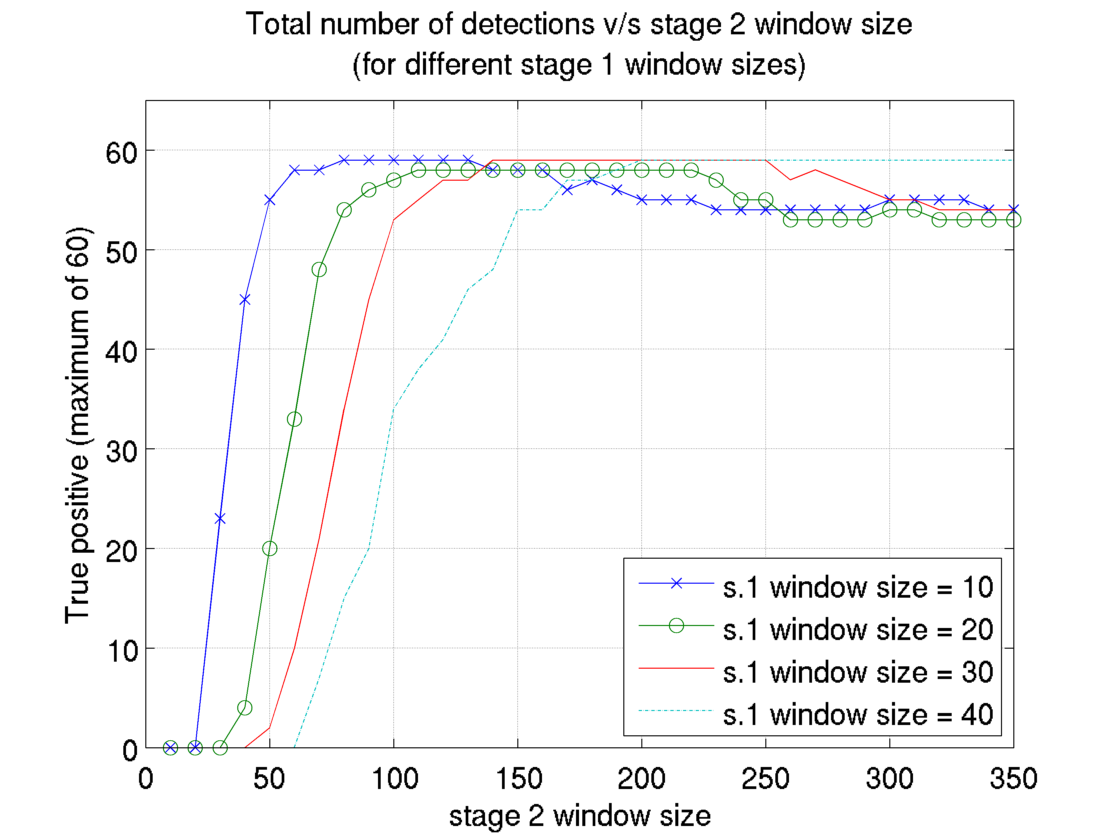}
     \caption{The increase in window size beyond a certain point reduces the number of detections by a small number.}
   \end{subfigure}
   \\
   \begin{subfigure}{1\linewidth}
   \centering
     \includegraphics[scale=0.55]{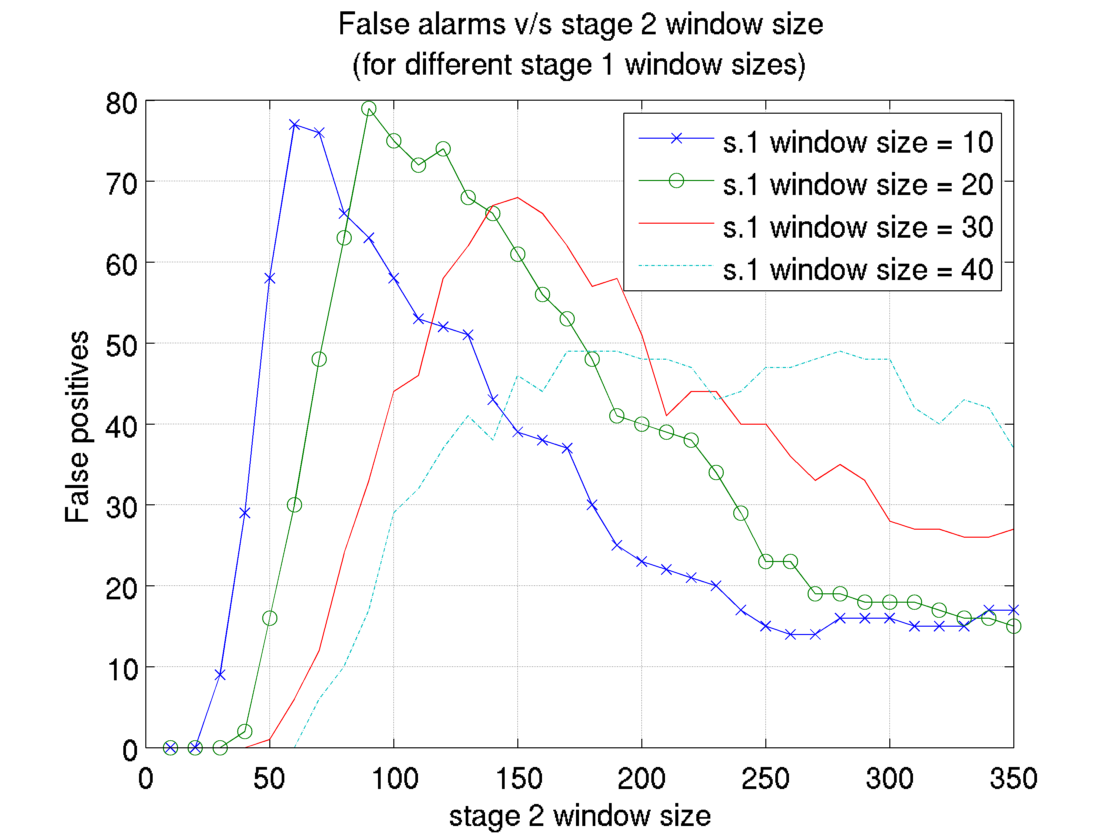}
     \caption{False alarms are reduced the larger the stage-2 window size is.}
   \end{subfigure}
 \caption{At the second stage, a larger window size tends to have a better performance overall.}
  \label{fig:det-false-p2}
\end{figure}

The impact of changing the stage-2 window size on the number of true and false positives is shown in figure \ref{fig:det-false-p2}. Concerning the number of detections, increasing the size of the stage-2 window shows an increase in these detections, which tends to settle (with a slight dip) beyond a particular point (around 125 samples). At the same time, the number of false alarms increases sharply before it peaks and decreases with an increasing window. This validates the conjecture made before: increasing the stage-2 window size allows us to have a better detection methodology, as it decreases the number of false alarms, while not diminishing the true positives by much.

\begin{figure}
  \centering
    \includegraphics[width=0.55\textwidth]{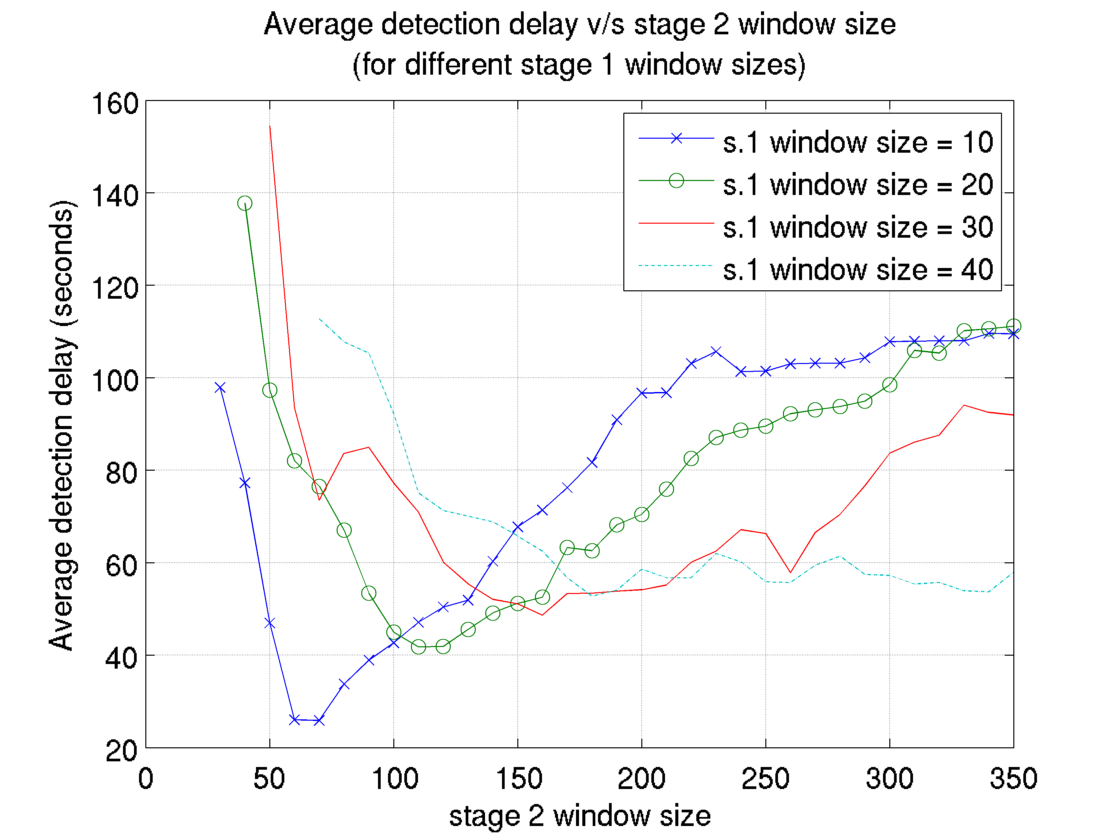}
  \caption{The stage-2 window size increase leads to a decrease in false alarms, but that comes at the expense of an increase in detection delay.}
  \label{fig:delay}
\end{figure}

However, and as expected, increasing the stage-2 window size has a direct effect on delays. This is readily observed in figure \ref{fig:delay}. Though the delay starts by being elevated with short window sizes, this can be explained by the fact that the number of leaks detected during with that configuration is very low, and the data is therefore not representative. However, once the delay reaches a minimum value, it starts to climb back up with increased stage-2 window size. Indeed, the confidence in the detection increases, but the consequence is that the algorithm has to wait for more and more samples before making the decision.\\

\begin{figure}
\centering
   \begin{subfigure}{1\linewidth}
   \centering
    \includegraphics[scale=0.55]{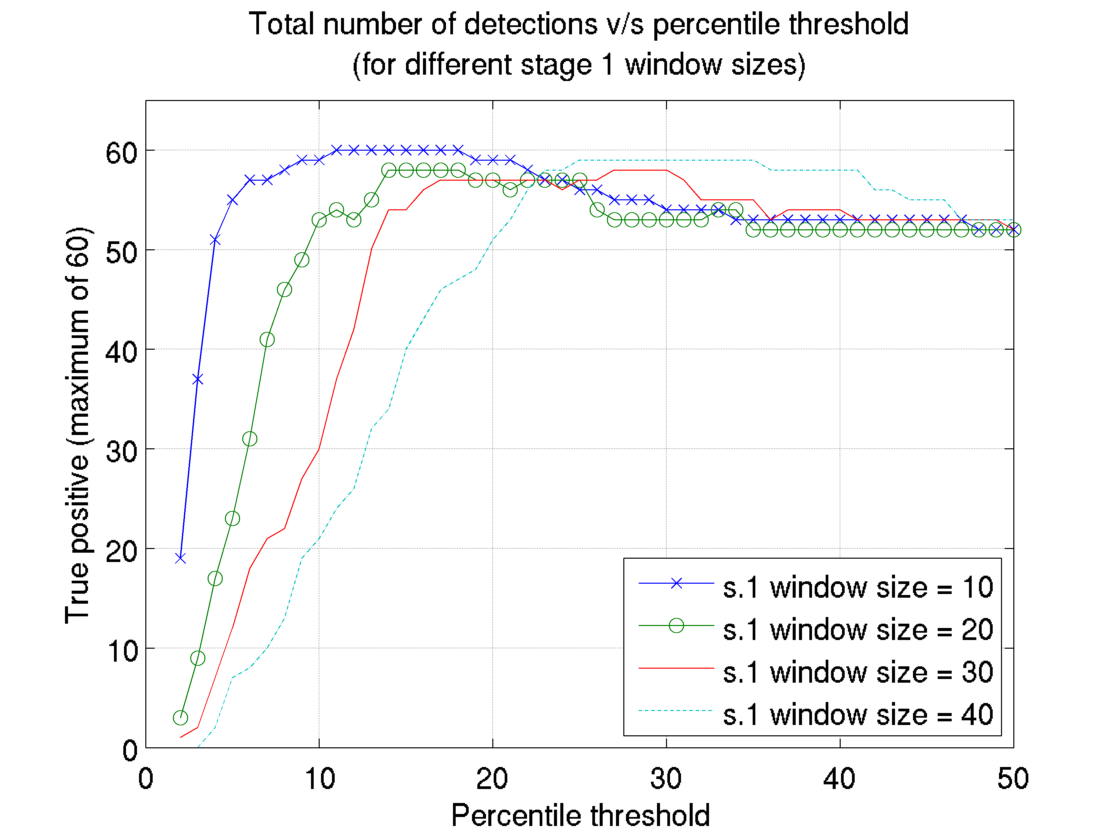}
     \caption{A percentile threshold in the 10-20\% seems to lead to a higher number of detections.}
   \end{subfigure}
   \\
   \begin{subfigure}{1\linewidth}
   \centering
     \includegraphics[scale=0.55]{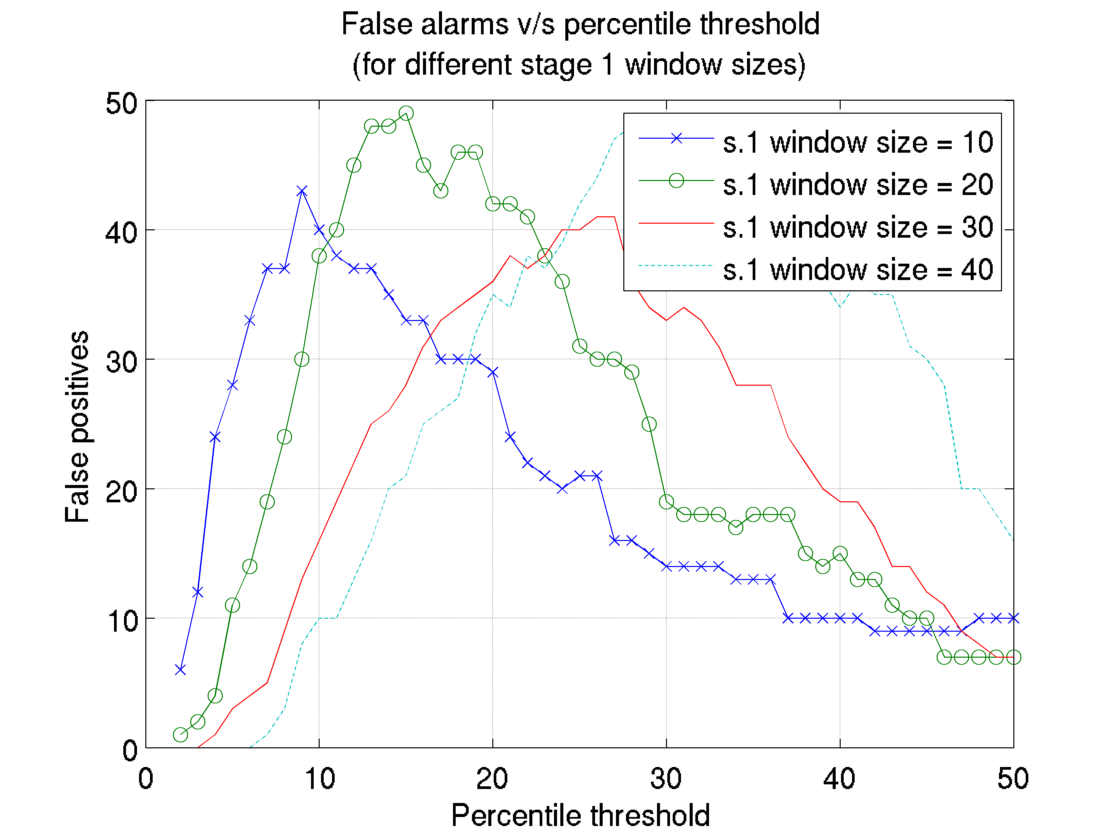}
     \caption{The number of false alarms rises quickly with the percentile, before it rolls off at a slower rate.}
   \end{subfigure}
 \caption{The percentile threshold parameter is very important in determining the performance of these routines.}
  \label{fig:perc}
\end{figure}

As explained before, the percentile parameter is an important one that determines the performance of our detection method. Looking at figure \ref{fig:perc}, we notice that increasing this parameter has a similar effect to increasing the stage-2 window size: the number of detections rises quickly before starting to dip slightly with increase percentile threshold, while the number of false alarms rises quickly in the beginning, before dropping as the percentile is increased.\\

\begin{figure}
\centering
   \begin{subfigure}{1\linewidth}
   \centering
    \includegraphics[scale=0.55]{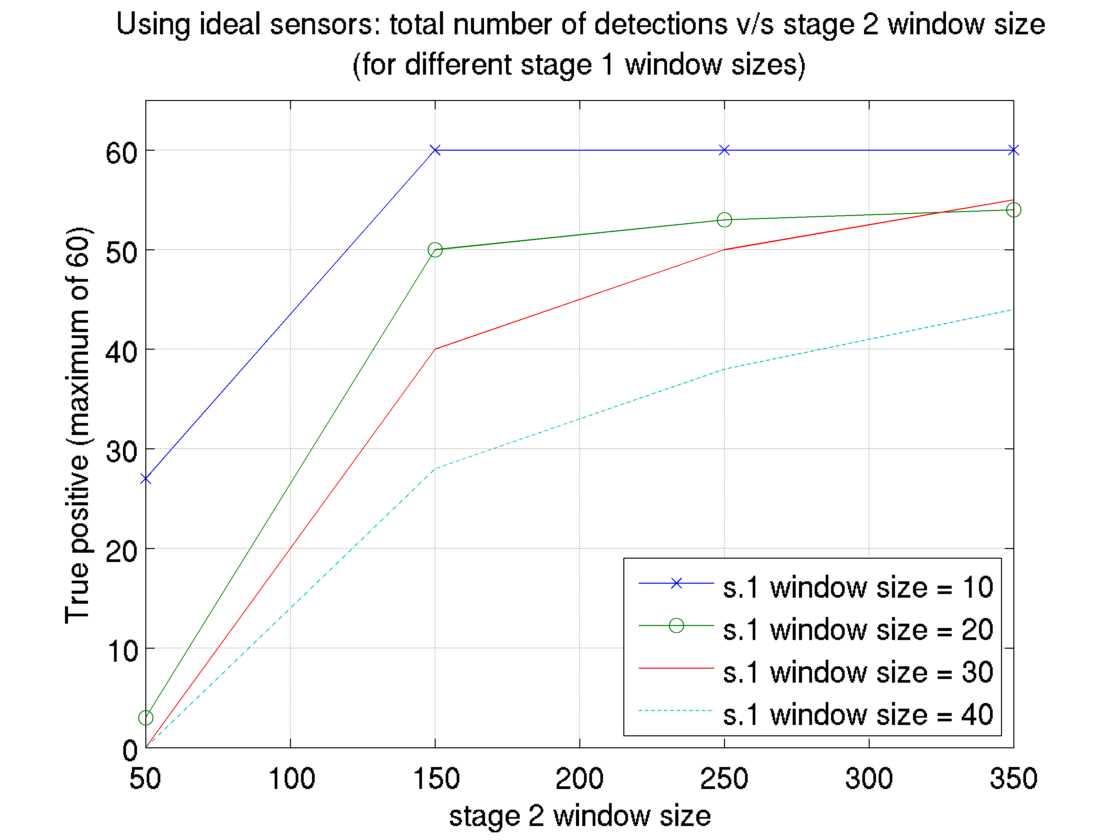}
     \caption{In some configurations, detecting all of the releases becomes possible.}
   \end{subfigure}
   \\
   \begin{subfigure}{1\linewidth}
   \centering
     \includegraphics[scale=0.55]{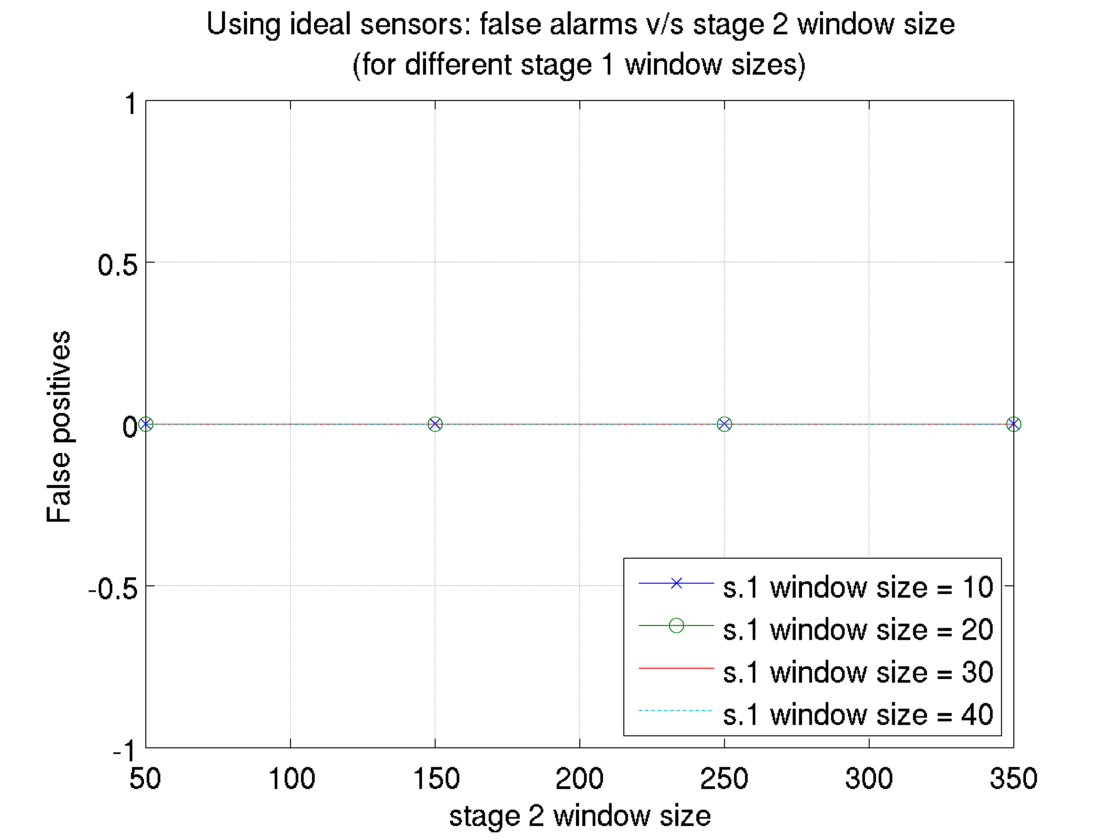}
     \caption{The rate of false alarms is zero across most configurations.}
   \end{subfigure}
 \caption{Given ideal sensors (with enhanced SNR), the detection problem would become much easier. This plot shows the result of the two stage detection algorithm presented here applied to ``ideal'' data generated from the experimental one.}
  \label{fig:ideal}
\end{figure}

At this point, it is worth noting that the most balanced configuration of this algorithm returned a detection rate of 55/60, with 7 false alarms, and an average delay of 108 seconds. Though these results seem promising, they leave a lot of room for improvement. The detection rate is satisfactory, especially when compared to the absence of widespread reliable detection methods on the market today, but a 100\% rate would be desirable of course. A delay of more than 100s, could represent some challenges for a refinery workers in responding to an alarm, so reducing it to below 1 minute is also desired. Finally, a false alarm rate of 7 over a period of three days seems excessive, even if these alarms were short-lived. Still, a false positive rate of 1 per plant per month (or per year) would be highly desirable, especially with an elevated detection rate to accompany it. To reach these desirable results, we would require an improvement in sensor technology of at least one order of magnitude in SNR. Concerning the lifetime of the sensors, and to enable a 10-year deployment without the need to change batteries, a reduction of power consumption of two orders of magnitude is required for reliable detection.\\
With this in mind, we ``massage'' our experimental data to reduce its noise as proposed and run the algorithm again. The results are shown in figure \ref{fig:ideal}. It is clear that certain configurations of our detection methodology would give us a 100\% detection rate (while others give a slightly reduced rate). Most importantly however, is the false alarm rate, which stays at zero across all configurations considered here.

\subsection{Localization Results}
\begin{figure}
  \centering
    \includegraphics[width=0.55\textwidth]{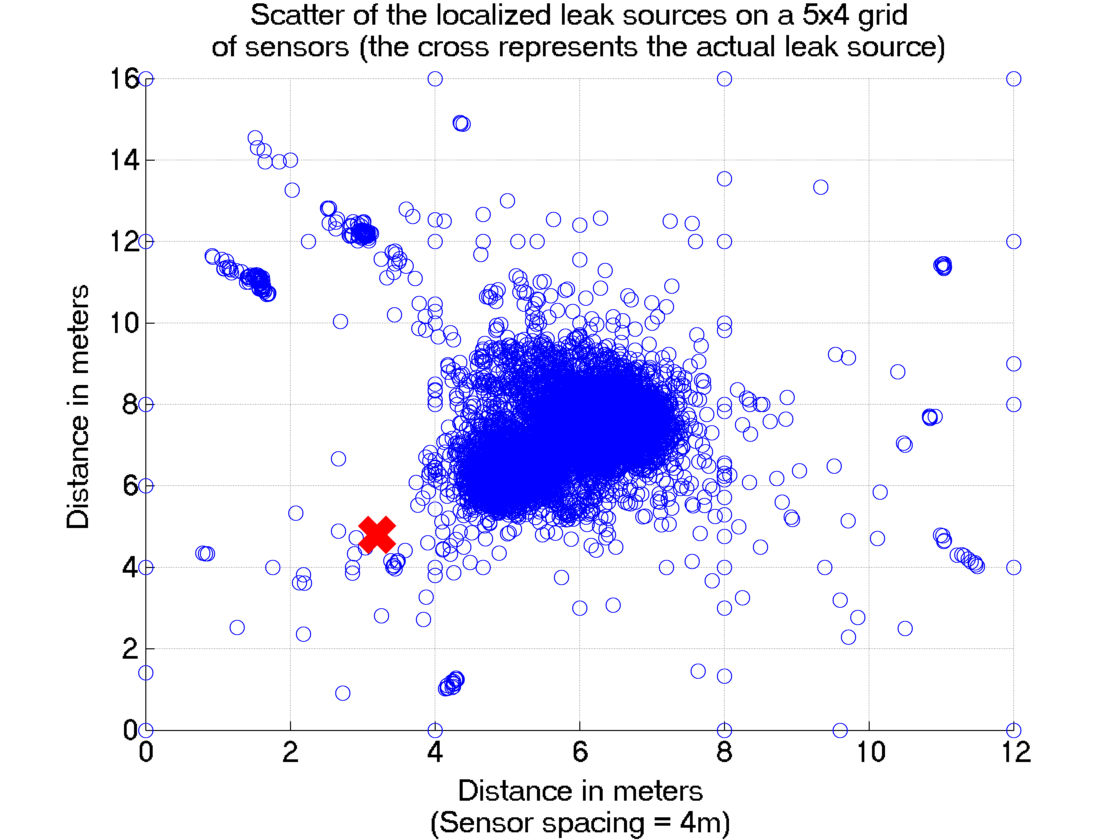}
  \caption{Scatter plot of all of the detections (during many different configurations). The offset seen between the real source of the leak and the conglomeration of detections could be explained by the fact that no sensors were present directly above the source.}
  \label{fig:scatter}
\end{figure}

\begin{figure}
  \centering
    \includegraphics[width=0.55\textwidth]{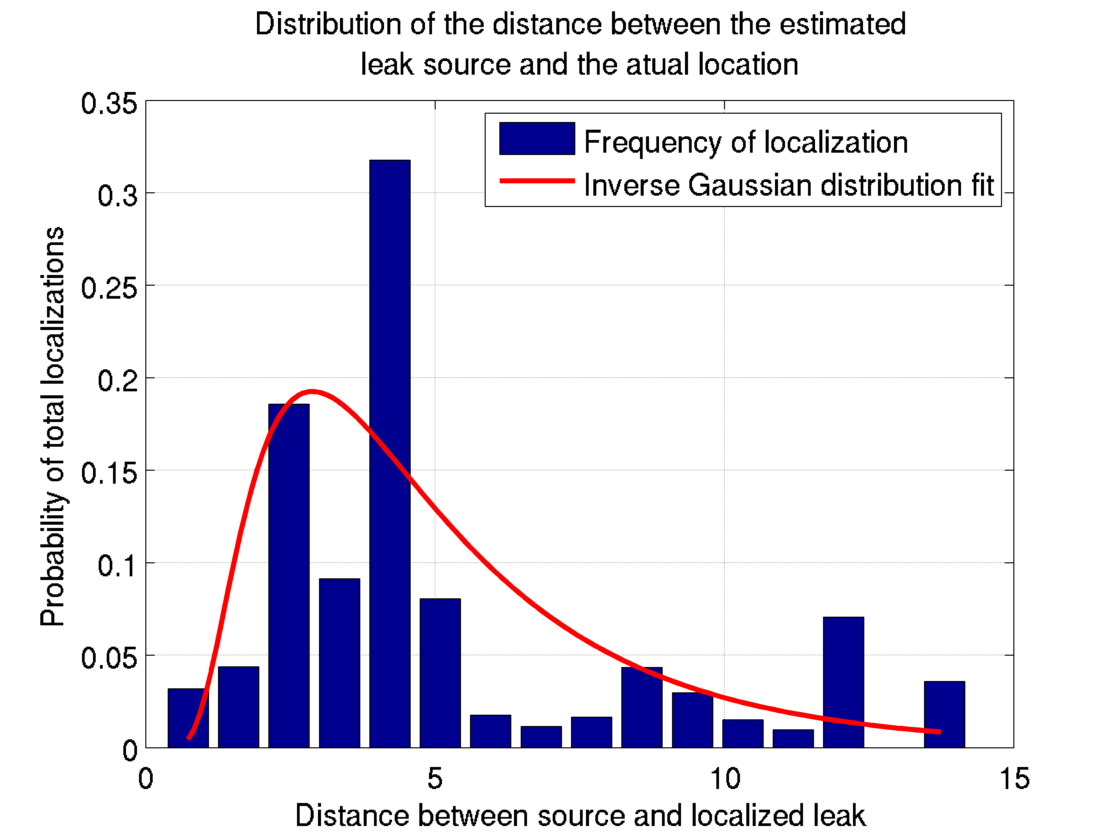}
  \caption{A histogram of the distances between the detections and the real source is leak is shown in this figure. More than 50\% of all detections are within 3 meters of the source.}
  \label{fig:loc-dist}
\end{figure}

\begin{figure}
    \includegraphics[width=0.55\textwidth]{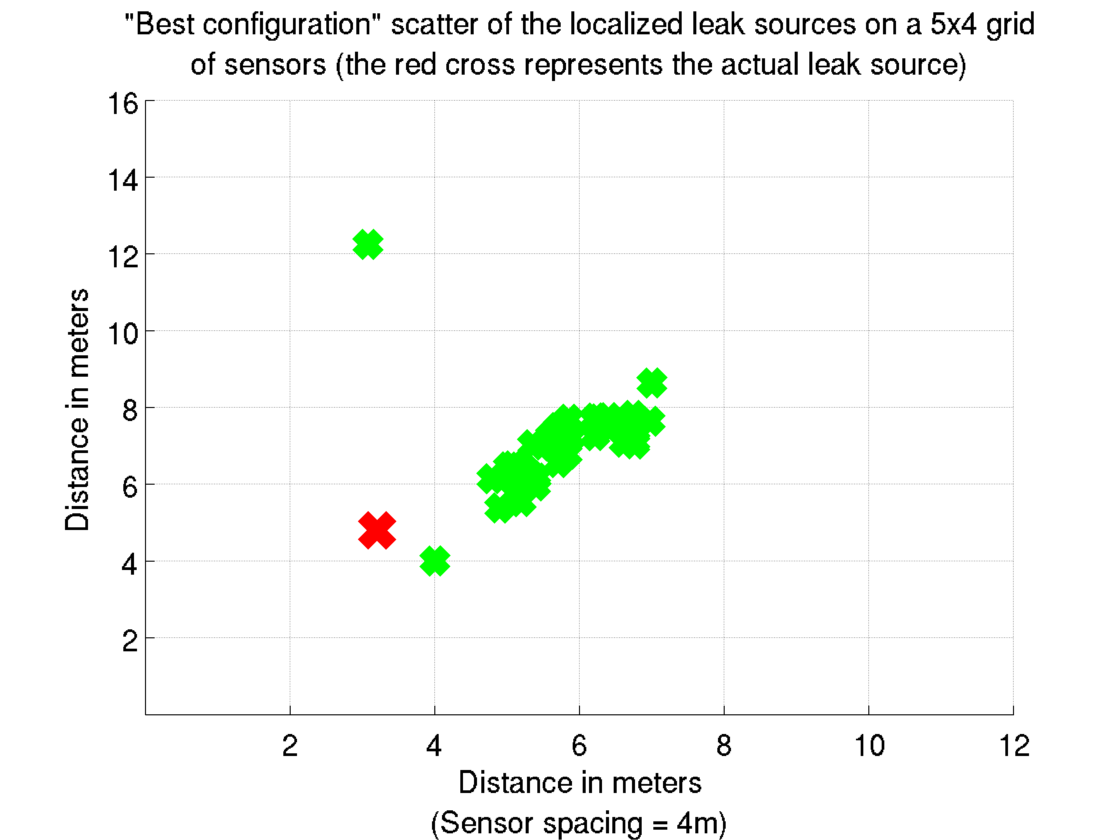}
    \caption{In the preferred configuration (resulting in 55 detections out of 60, with 7 false alarms), the localization results appear even closer to the actual source of the leaks.}
  \label{fig:loc-best}
\end{figure}

\begin{figure}
\centering
   \begin{subfigure}{1\linewidth}
   \centering
    \includegraphics[scale=0.55]{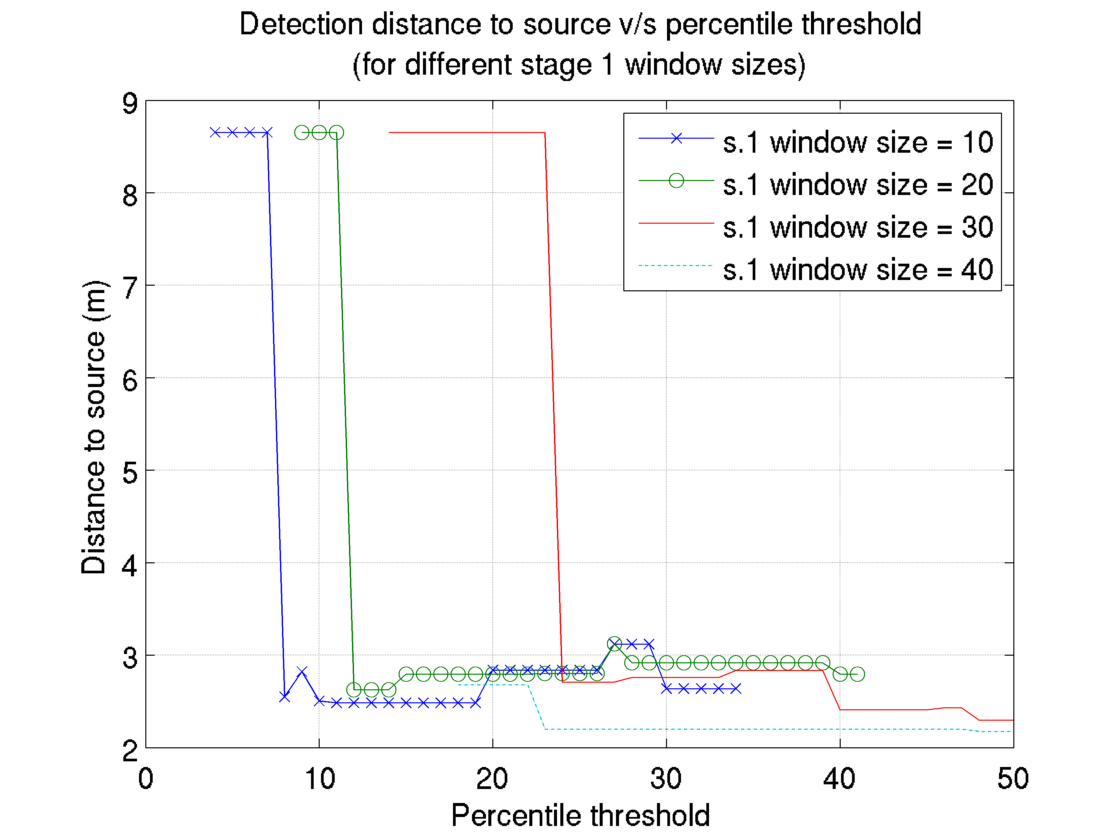}
     \caption{Larger percentile thresholds tend to yield better localization results since the response of more sensors is taken into account.}
   \end{subfigure}
   \\
   \begin{subfigure}{1\linewidth}
   \centering
     \includegraphics[scale=0.55]{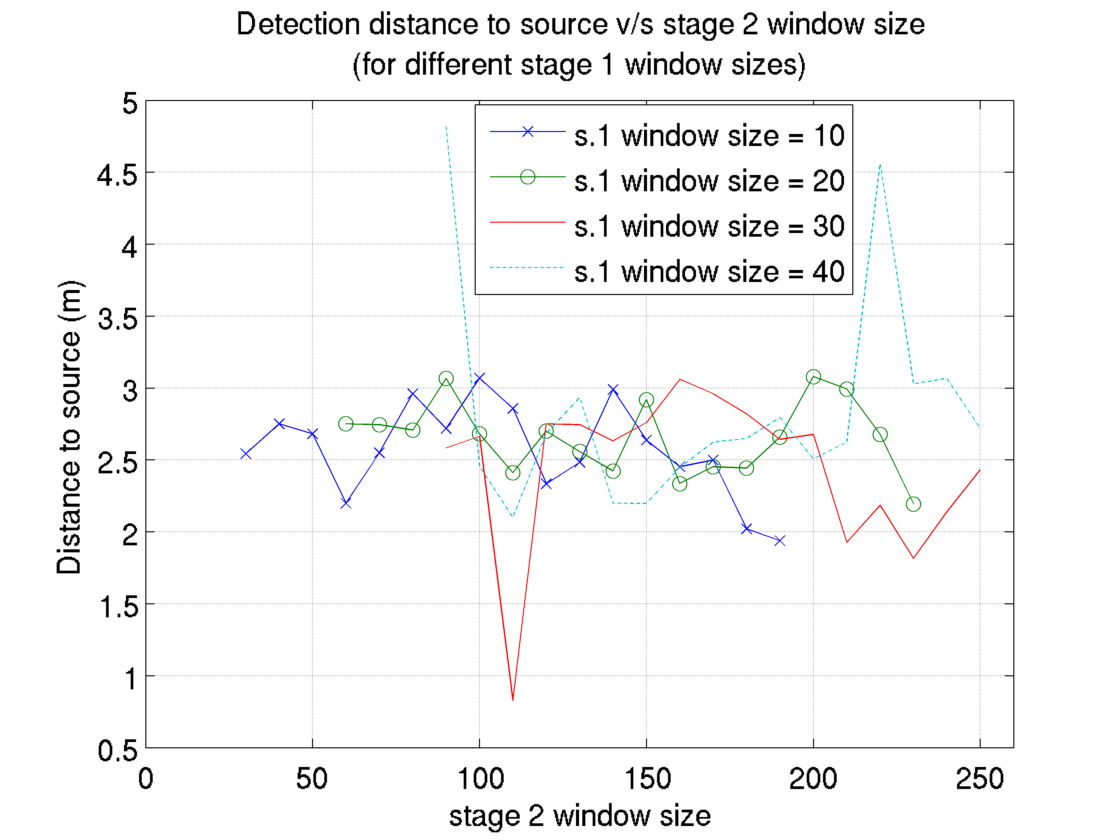}
     \caption{Stage-1 and stage-2 window sizes largely do not affect the localization.}
   \end{subfigure}
 \caption{Distance between the detected leak and the real source, for on particular release under various algorithm configurations.}
  \label{fig:loc-params}
\end{figure}

We now look at the localization results given our noisy sensors covering the test site. Upon finding a detection, the center of mass routine is called upon the concentration data, and the point of highest mean in both the X and Y directions is identified as the leak source estimate. Figure \ref{fig:scatter} shows a scatter of these detection estimates (for the different parameters described above) along with the true source of the leaks. Most of the detections seem clustered in the middle of the sensor grid, but with a slight skew towards the actual source. One potential explanation to this phenomenon is the absence of sensors directly above the leak release point. This means that the algorithm is converging towards the closest sensor(s) in the vicinity of this source.\\
Looking at the distribution of distance between the estimated leak source and the actual one in figure \ref{fig:loc-dist}, we can see that it follows an inverse Gaussian trend. Most detections actually occurred less than three meters away from the leak source, and a striking majority were localized less than 5 meters away. Also shown in figure \ref{fig:loc-best}, are the localizations of the 55 ``best'' detections which were obtained as described above (with the 7 false alarms and the delay of 108 seconds). Finally, figure \ref{fig:loc-params} shows the distance between the detected leak and the actual source, for one particular leak and across different configurations. Larger percentile thresholds lead to better detections since the concentration measurements of more sensors is taken into account. In general varying the stage-1 and stage-2 window sizes does not affect the localization results greatly. The main reason is that the localization routine is purely a center of mass approach, and should not be affected by the parameters. Hence, unless the detection is distant, localization results will be more or less independent of configuration parameters. However, by looking at particular leaks, some configurations were identified as problematic across various gas releases. The results shown confirm the reliability of the localization method, with no estimates found in unusual locations (near the edge of the network for example). In a realistic deployment, the localization figures presented here would be very helpful for workers who are familiar with the equipment present in the vicinity of the sensors, and who should quickly be able to identify the source of the leak.\\

\section{Conclusions}
\label{sec:conclusion}
In this study, we proposed, implemented and validated a wireless distributed gas leak solution for industrial places. In our system architecture, many gas sensors are placed around a region of interest in a plant, and they all report to a single location. These sensors are duty-cycled in time and space to conserve energy (as they are battery powered). At the gateway, an algorithm is run on the measured concentration data which allows the detection and localization of the leaks. In our experiment (with 20 sensors and a monitoring area of $200m^2$), we were able to detect 55 out of the 60 releases, with an average delay of 108 seconds, and a localization accuracy under 5m. Over the period of three days, we saw 7 false alarms. Though we can achieve these results today, there is still some work to be done to get this idea into productization. In this study we made an attempt at being as agnostic to the leak source as possible. Studying the relationship between flow rates and nozzle sizes would be beneficial and is left for future research. As the wireless communication reaches new boundaries in reliability, and as efficient microcontrollers become cheaper and less power hungry, the only component left to be improved is the sensor. Certainly, improvements on the detection algorithm could help reduce the false alarm rate and increase the detection rate, but the sensing hardware would be a better place to start. An order of magnitude improvement in signal-to-noise needs to be accomplished in explosive gas sensors (to reduce false alarms). A faster wake-up and quicker response time is required for faster detections. Finally, two orders of magnitude of improvement in energy consumption is needed to extend the lifetime of the device to 5 years. As a concluding note, and though this study was done with industrial plants in mind, we feel that similar approaches can be accomplished in cities of the future, where a gas detection and localization system can help address the problems of leaks in urban gas pipelines.

\section*{Acknowledgements} \label{sec:ack}
We would like to thank Chevron Energy Technology Company and Chevron Information Technology Company for funding and supporting this project. Many thanks also go to the Texas A\&M Engineering Extension Facility and their helpful staff, and the Texas A\&M team led by Dr. Yi Liu. Finally, we would like to acknowledge Beverly Coleman and Julio Cedeno for driving crucial change in today's corporate world.

\end{document}